\documentclass[11pt]{article}

\usepackage{graphicx}
\usepackage{cite}
\usepackage{dcolumn}
\usepackage{bm}
\usepackage{amssymb}
\usepackage{amsmath}
\usepackage{epsfig}    
\usepackage[colorlinks=true]{hyperref}
\def\beq{\begin{equation}}
\def\eeq{\end{equation}}
\def\ba{\begin{array}}
\def\ea{\end{array}}
\def\bea{\begin{eqnarray}}
\def\eea{\end{eqnarray}}

\def\sq2{\sqrt{2}}
\def\sqs{\sqrt{s}}
\def\ab1{{\rm ab}^{-1}}
\def\End{\end{document}}

\newcommand{\gm}{\gamma^\mu}
\newcommand{\smn}{\sigma^{\mu \nu}}
\newcommand{\Wmn}{W_{\mu \nu}}
\newcommand{\Bmn}{B_{\mu \nu}}
\newcommand{\sqd}{\sqrt{2}}

\renewcommand{\P}{\ensuremath{\mathcal{P}_e}}

\newcommand{\veps}{\varepsilon}
\renewcommand{\bar}[1]{\overline{#1}}

\renewcommand{\P}{\mathcal{P}}
\newcommand{\vp}{\varphi}

\newcommand{\C}{\overline{C}}
\newcommand{\F}{\overline{F}}

\newcommand{\e}[2]{&\ensuremath{
  \begin{array}{r}
    #1\\[-2pt]#2
  \end{array}}}

\begin{document}

\title{Single top production at linear $e^-e^+$ colliders}

\author{
Angel Escamilla, \; Antonio O.\ Bouzas,\thanks{
abouzas@fis.mda.cinvestav.mx, corresponding author.}
\; F. Larios\\
Departamento de F\'{\i}sica Aplicada,
CINVESTAV-M\'erida, A.P. 73, 97310 M\'erida,
Yucat\'an, M\'exico}

\def\thisday{dec 07, 2017} 


\maketitle

\begin{abstract}
We study single top production at linear lepton colliders with
$\sqs = 0.5$, 1 and 3 TeV.  A preliminary analysis shows that
despite the large $t\bar t$ background at $\sqs = 0.5$
and 1 TeV it is possible to obtain a good sensitivity to the
$tbW$ vertex; even more so at 3 TeV when single top becomes
the dominant mode of production.  Concerning the four dimension
six operators involved, two of them simultaneously generate
$ttZ$ couplings and their sensitivity decreases with energy.
The opposite is true for the other two operators.  Single top
production at these machines is also useful to probe charged-current 
four fermion operators $e\nu bt$, some of which are
related to $tbW$ operators through the equations of motion.
\end{abstract}

\section{Introduction}

Future linear lepton colliders such as the International Linear
Collider (ILC) \cite{ilc1,ilc2,ilc3a,ilc3b,ilc4} and the Compact
Linear Collider (CLIC) \cite{clic2,clic3} have the top quark as one of
their main areas of research.  In particular, an extensive effort based
on the dimension 6 operators \cite{grz10} of the Standard Model
Effective Field Theory is being developed for the top-quark physics
program at these colliders as well as at the LHC \cite{threviews}.
For the LHC, significant limits have been obtained when the effective
couplings enter in loop-level \cite {blas,zha14,loopbounds} as well as
tree-level processes \cite{maltoni15, coimbra}, and many experimental
measurements can be found in the literature, for $t \bar t$
production, $W$-helicity in top quark decay, rare top decays,
same-sign tops production, single-top, mono-top and multiple-top
production \cite{4tops}.

In the context of top-quark production in $e^-e^+$ colliders, so far
most of the interest has been placed on $t\bar t$ and $t\bar t H$
production at the ILC for beam energies of $\sqrt{s}=0.5$ and 1 TeV
\cite{eett}, and very few studies have been done on single-top
production \cite{boo96,boo01,penu11,schwinn,fuster}.  It is now known
that $ttZ(\gamma)$ couplings will be far better probed in this machine
than at the LHC \cite{agsa12,englert17}.  However, with respect to the
$tbW$ coupling the LHC is already providing very strong limits through
single-top production and $W$-helicity in top decays \cite{cms17}. In
contrast, $t\bar t$ production at the linear collider has very little
sensitivity to $tbW$ even for angular distributions of decay products
\cite{kolodziej}.  To date, there is no study on the potential of
single-top production in $e^-e^+$ collisions to probe the effective
$tbW$ coupling.  This is one of the goals of this paper: to find out
what is the sensitivity to this coupling and how it compares with the
potential of $e^+e^- \to t\bar t$ as well as the LHC.  Furthermore, we
also study the sensitivity to four-fermion operators $e\nu bt$
relevant to this process.

We refer in this study to the basis of dimension-6,
SU(3)$\times$SU(2)$\times$U(1) gauge-invariant operators provided in
\cite{grz10}.  In tables \ref{tab:ilcops} and \ref{tab:ilcops4} we
show the flavor-diagonal operators in that basis that are relevant to
top quark production at the ILC and CLIC.  The notation used here is
standard: $\tau^I$ are the Pauli matrices, $\varphi$ is the SM Higgs
doublet with $\tilde \varphi = i\tau^2 \varphi^*$, $q_L$ is the third
generation left-handed $SU(2)$ doublet, $t_R$ and $b_R$ are the
right-handed $SU(2)$ singlets.  The covariant derivative is defined as
$D_\mu \varphi = \partial_\mu \varphi - i g/2 \tau^I W^I_\mu \varphi -
i g'/2 B_\mu \varphi$.  As described in more detail below, we will
follow the operator normalization used in \cite{zha14,maltoni15}.


\begin{table}[ht!]
  \centering
  \begin{tabular}{|c|c|c|}\hline
    $tt Z+tt A$ & $ttZ+tt A+ tbW$ &  $tbW$ 
    \\\hline
    $O_{\varphi q}^{(1)33}=
  \varphi^\dagger i \overline D_\mu \varphi \bar q_{L} \gm q_L$ &
    $O_{\varphi q}^{(3)33}=
 \varphi^\dagger i \overline D^I_\mu \varphi \bar q_L \tau^I \gm q_L$ &
     $O_{\varphi ud}^{33}=
  \tilde  \varphi^\dagger i D_\mu \varphi \bar t_R \gm b_R$ 
    \\\hline
    $O_{\varphi u}^{33}=
   \varphi^\dagger i \overline D_\mu \varphi \bar t_R \gm t_R$ &
     $O_{uW}^{33}=
  \bar q_L \smn t_R \tau^I \tilde \varphi \Wmn^I$ &
      $O_{dW}^{33}=
  \bar q_L \smn b_R \tau^I \varphi \Wmn^I$ 
    \\\hline
    $O_{uB}^{33}=
    \bar q_L \smn t_R \tilde \varphi \Bmn$ &
    -- &
    -- 
    \\\hline
  \end{tabular}                               
  \caption{Operators relevant for Top production at ILC.
    Top-gauge Boson.  Indices $33$ stand for third generation
  quarks.}
  \label{tab:ilcops}
\end{table}


\begin{table}[ht!]
  \centering
  \begin{tabular}{|c|c|c|}\hline
    $eett$ & $eett+e\nu tb$ &  $e\nu tb$ 
    \\\hline
    $O_{\ell q}^{(1)13}=
    \bar \ell_L \gamma_\mu \ell_L \bar q_L \gm q_L$ &
       $O_{\ell q}^{(3)13}=
 \bar \ell_L \gamma_\mu \tau^I \ell_L \bar q_L \gm \tau^I q_L$ &
       $O^{13}_{\ell edq}=
    \bar \ell_L e_R \bar b_R q_L$ 
   \\\hline
    $O^{13}_{eu}=
    \bar e_R \gamma_\mu e_R \bar t_R \gm t_R$ &
       $O_{\ell equ}^{(1)13}=
    \bar \ell^j_L e \; \epsilon_{jk} \; \bar q_L^k t$ &
      --
      \\\hline
    $O^{13}_{\ell u}=
  \bar \ell_L \gamma_\mu \ell_L \bar t_R \gm t_R$ &
$O_{\ell equ}^{(3)13}=
  \bar \ell_L^j \sigma_{\mu \nu} e_R \; \epsilon_{jk}\;
  \bar q_L^k \smn t_R$ &
    --
     \\\hline
$O^{13}_{qe}=
    \bar q_L \gamma_\mu q_L \bar e_R \gm e_R$ &
    -- &  -- 
    \\\hline
  \end{tabular}                               
  \caption{Four-fermion operators relevant for Top production
    at ILC.  Indices $13$ stand for first family leptons and
    third family quarks.}
  \label{tab:ilcops4}
\end{table}


The basis operators generating couplings of the top quark to the gauge
bosons are displayed in table \ref{tab:ilcops}.  As shown there, there
are three operators that only generate neutral current (NC) $ttZ$ and
$tt\gamma$ vertices, and two that generate both charged-current (CC)
and NC couplings.  For these five operators the ILC $t\bar t$ process
can indeed surpass the potential of the LHC \cite{cao15,englert17}.
The remaining two operators in table \ref{tab:ilcops} generate solely
CC $tbW$ effective couplings that cannot be sensitively probed by
top-pair production.  In this paper we discuss whether the single-top
production mode at the linear colliders would be able to give bounds
for the two purely CC two operators similar to or more stringent than
the LHC, and how the limits on the two mixed NC/CC operators from
single-top compare to those from top-pair production.

We parenthetically point out here that, strictly speaking, the
distinction between NC operators and CC operators that are sensitive
to the top-pair and the single-top processes separately is not fully
clear-cut.  Indeed, it has been pointed out that off-shell effects in
top pair production can indeed bring sensitivity to the $tbW$
coupling, and in particular be used to measure the top-quark width
with great accuracy \cite{tait06,widthteams}, which argues in favor of
the notion that the potential of the ILC and CLIC machines in studying
top-quark physics will go beyond the context of on-shell $t\bar t$
production.

As has been pointed out in \cite{bach12}, a consistent analysis of
top-gauge boson operators cannot exclude the effects of four-fermion
operators.  Indeed, the choice of dimension 6 basis top-gauge boson
operators implies that other operators of the same type are deemed
redundant because of the equations of motion \cite{grz10}.  These
equations of motion involve four-fermion operators that are chosen to
appear in the list of independent operators and, therefore, must be
included in the analysis if it is to be mathematically consistent and
model independent.

The basis operators generating four-fermion vertices involving the top
quark are shown in table \ref{tab:ilcops4}.  As with the quark-gauge
boson operators we focus on operators containing only third-generation
quarks and, in the case of four-fermion operators, first-family
leptons.  As seen in table \ref{tab:ilcops4},
there are four operators generating purely $eett$ vertices that are
related only to NC $t\bar t$ production.  We will not consider them in
this paper; a recent study on the ILC potential to probe them can be
found in \cite{rindani}.  Another goal of this paper is to obtain the
limits set by single-top production on the remaining four operators in
the table, which generate CC-type $e\nu tb$ couplings relevant to that
process. 

This article is organized as follows.  In section \ref{sec:topatee} we
discuss in detail the SM process of single-top production and decay at
an $e^- e^+$ collider, as well as its reducible and irreducible
backgrounds and the role of beam polarizations. In
section~\ref{sec:singletoperators} we review the flavor-diagonal
effective operators relevant to single-top production and discuss the
recent LHC results on effective top-gauge boson couplings and the
projected sensitivity of top-pair production at the ILC to those
couplings, which set the context against which single-top production
at ILC and CLIC must be analyzed.  In section \ref{sec:res} we obtain
bounds on the effective couplings from the single-top total cross
section, both at the individual-coupling level and for pairs of
couplings, at $\sqs=0.5$, 1 and 3 TeV for certain ranges of
experimental uncertainties at each energy.  Finally, in
section~\ref{conclusions} we present our conclusions.

\section{Top quark production at an $e^+ e^-$ collider.}
\label{sec:topatee}

To better understand why we have chosen the single top process defined
below in (\ref{eq:prddcy}) let us review the context of top pair and
single top production in an $e^+ e^-$ collider.  Unlike the LHC,
$t\bar t$ production at the ILC is generated by the electroweak
interaction and becomes an irreducible background for single top
production.  Single top production can be hard to distinguish from
$e^+ e^- \to t\bar t$ particularly near the threshold region.  This
intermingling makes off-shell effects in top pair production sensitive
to the $tbW$ vertex. Therefore, it can be used to probe the $tbW$
vertex and the top width \cite{tait06,widthteams}.  At tree level
$\sigma(e^+ e^- \to t\bar t)$ is given by just two diagrams
($s$-channel $Z$ and $\gamma$). At $\sqs =0.5$ TeV the cross section
is about 550 fb with an increase of about $15\%$ when QCD corrections
are included \cite{dekkers16}.  If we require one of the top quark
lines to be about $20$ GeV away from the resonance so as to obtain
single top events, the contribution from these 2 diagrams yields:
$\sigma(e^+ e^- \to tt^* \to t\bar b W^- + \bar t b W^+) \simeq 20$fb.
This does not mean that $tt^*$ is the main source of single top
production.  If we consider the $t\bar b W^-$ final state, with no CKM
mixing, we will find that there are a total of 7 diagrams, with only
two of them corresponding to $tt^*$, and that the cross section is
actually $\sigma(e^+ e^- \to t\bar b W^- + \bar t b W^+) \simeq 50$fb.
At this level, one can ask what are the possible decay channels and
the most interesting ones. We should bear in mind that the final
states coincide with the well known $t\bar t$ decay channels.  The
dileptonic channel, with the $4.5\%$ fraction of about $2.3$ fb would
yield about 2300 events with a luminosity of 1 $\ab1$ before cuts.
This is actually a very rough estimate, let's consider specifically
$e^+ e^- \to \bar b \mu^- \bar \nu_\mu b e^+ \nu_e$ that with a
$1.1\%$ fraction we would expect to contribute with about $0.55$fb.
It turns out that this process in particular has 438 diagrams, indeed
most of them with no $t$-lines.  After imposing a cut on the invariant
$M_{b\bar b}$ to be away from the $Z$ and Higgs boson resonances the
cross section reduces to just about $0.2$fb \cite{schwinn}.  The
dileptonic channel thus seems to yield rather poor statistics.  Let us
now consider the semileptonic mode, with final states ${\ell^- \bar
  \nu \bar b bjj}$ or ${\ell^+ \nu b \bar b jj}$.  Whether the lepton
is an electron or a muon we now expect to have about a $2\, \times
7.2\%$ fraction that is about $2\, \times 3.6$ fb for each possibility
$\ell = e,\mu$.  However, if $e^{\pm}$ is the lepton in the final
state, one desirable feature arises: $t$-channel diagrams appear.  In
$t$-channel diagrams there are no $ttZ(\gamma)$ vertices and the
sensivility goes only to the $tbW$ coupling.  From \cite{boo01} we
find that the actual cross section for $e^+ e^- \to t \bar b e^- \bar
\nu_e$ is about 3 fb where the invariant mass of the $b e^- \bar
\nu_e$ system is at least $20$GeV away from the top quark resonance.
With a $67\%$ fraction of the hadronic decay we then expect to have a
total of 4 fb for the semileptonic mode with the electron.  Kinematic
cuts will still reduce this number significantly, as we shall see
below, but yet enough cross section will survive that would yield good
statistics.  This is the final state of interest for this study: two
$b$-jets, two light quarks, an electron or positron and its neutrino.


\subsection{Semileptonic signal process and irreducible background}
\label{sec:sgnlirr}

The set of Feynman diagrams for top-quark production and decay in
$e^-e^+$ collisions in the SM in the semileptonic channel is a subset
of those for the six-fermion processes
\begin{equation}
  \label{eq:procs}
    e^-e^+ \rightarrow 
 q_u\overline{q}_d\, b\overline{b}\,e^-\overline{\nu}_e 
+
\overline{q}_u q_d\, b\overline{b}\,e^+\nu_e,
\quad \mathrm{with}\; q_u=u,c, \; q_d=d,s.
\end{equation}
The final states (\ref{eq:procs}) can be reached through two different
top production processes, one followed by hadronic top decay:
\begin{subequations}
    \label{eq:prddcy}
\begin{equation}
  \label{eq:hdrdcy}
  e^-e^+ \rightarrow
  \left\{
    \begin{array}{rr}
      t\overline{b}e^-\overline{\nu}_e, & t\rightarrow q_u
      \overline{q}_d b,\\
      \overline{t}be^+\nu_e, & \overline{t}\rightarrow \overline{q}_u
      q_d \overline{b},
    \end{array}
  \right.
\end{equation}
and the other one followed by leptonic decay:
\begin{equation}
  \label{eq:lptdcy}
  e^-e^+ \rightarrow
  \left\{
    \begin{array}{rr}
      t\overline{b}\overline{q}_u q_d, & t\rightarrow e^+\nu_e b,\\
      \overline{t}bq_u \overline{q}_d, & \overline{t}\rightarrow
      e^-\overline{\nu}_e \overline{b}.
    \end{array}
  \right.
\end{equation}
\end{subequations}
The process (\ref{eq:hdrdcy}) has been studied in \cite{boo96,boo01}
at the top-production level ($e^-e^+ \rightarrow
t\overline{b}e^-\overline{\nu}_e$).  Here, we extend that study to
include top decay and the process (\ref{eq:lptdcy}).

\begin{figure}[ht!]
\centering
\begin{picture}(420,210)
\put(0,0){\includegraphics[scale=0.9]{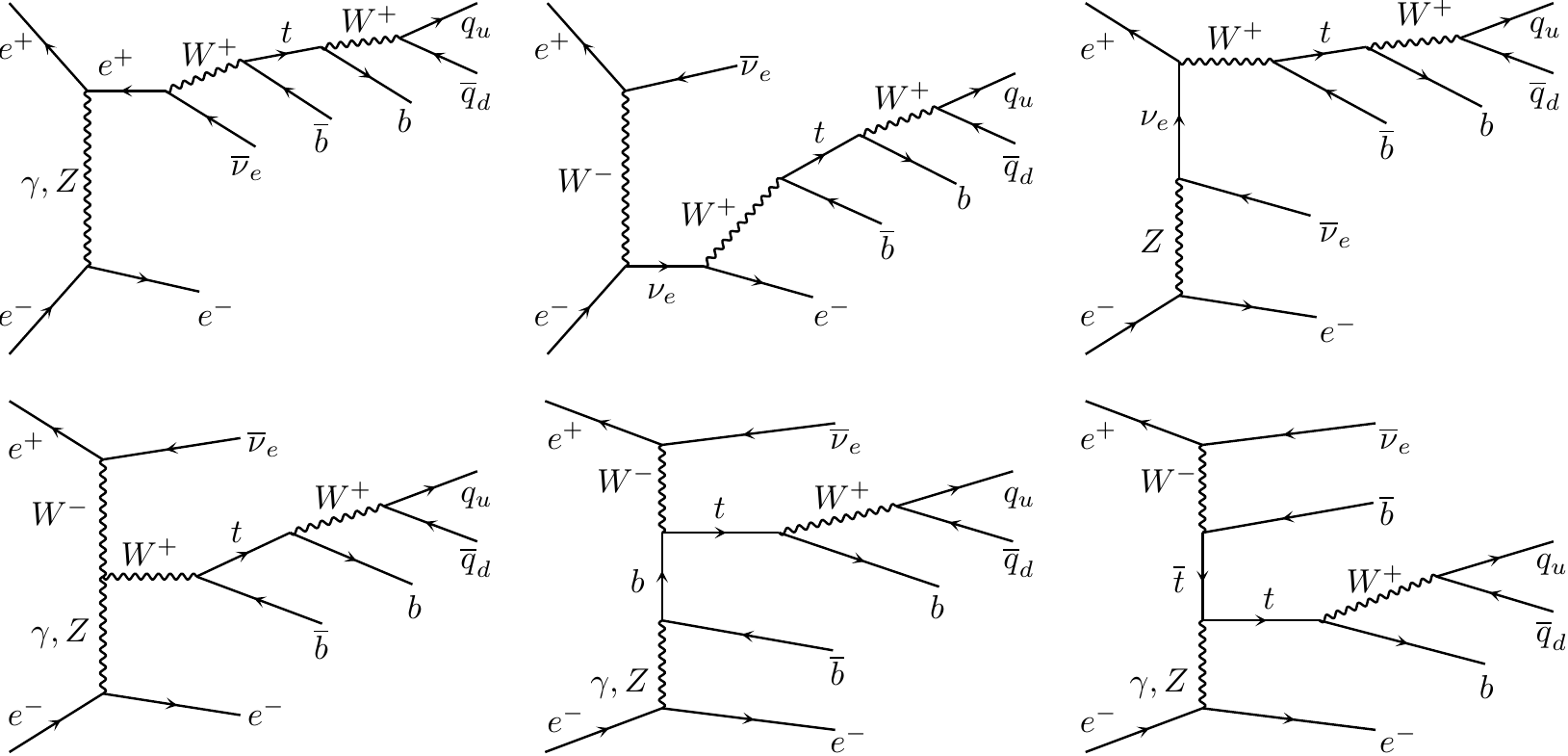}}
\end{picture}
\caption{Unitary gauge Feynman diagrams for single-top production in $e^-e^+$
  collisions with $t$-channel vector boson exchange.}  
  \label{fig:feyn1}
\end{figure}

\begin{figure}[ht!]
\centering
\begin{picture}(420,210)
\put(0,0){\includegraphics[scale=0.9]{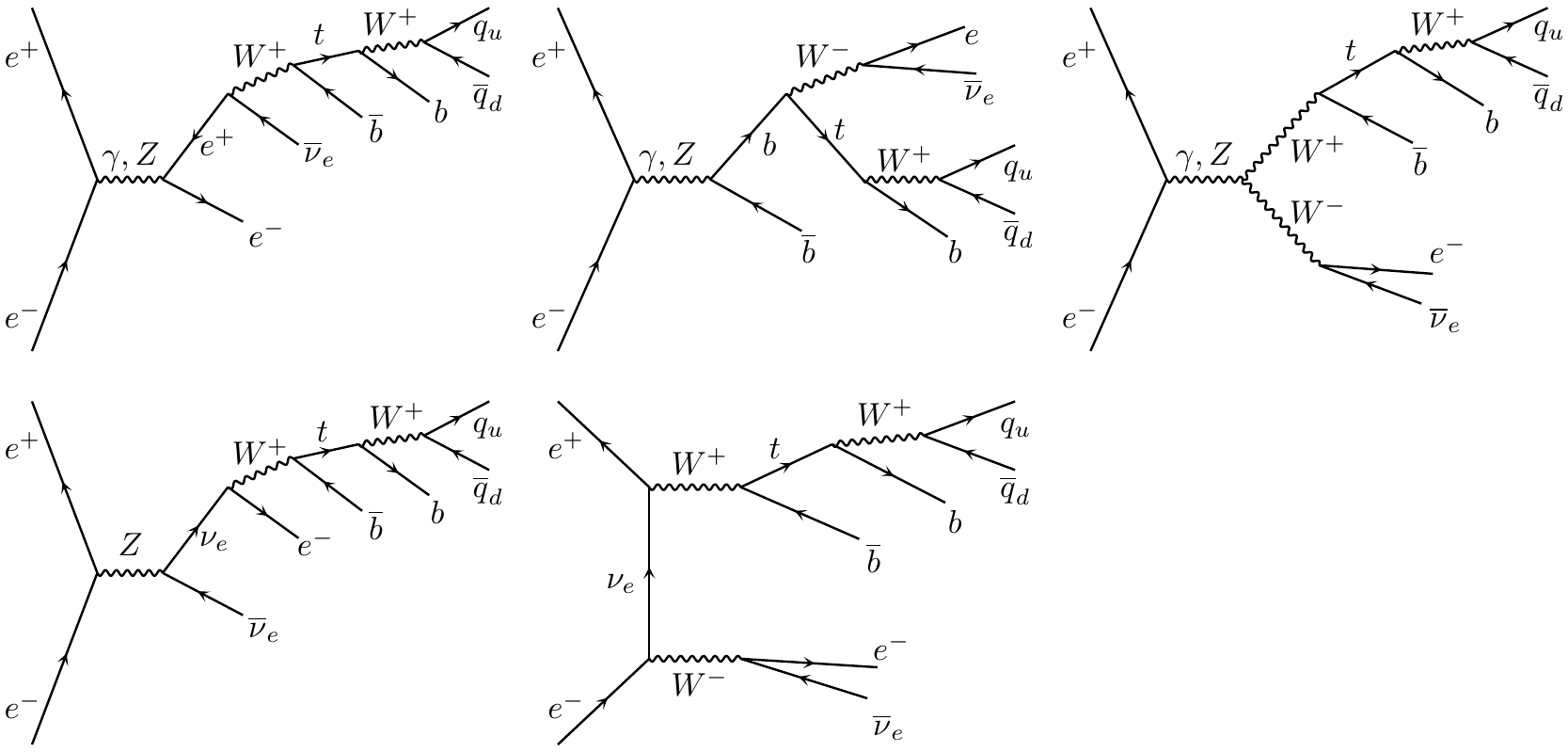}}
\end{picture}
\caption{Unitary gauge Feynman diagrams for single-top production in $e^-e^+$
  collisions with $s$-channel vector boson exchange and hadronic top decay.}  
  \label{fig:feyn2}
\end{figure}

\begin{figure}[ht!]
\centering
\begin{picture}(420,210)
\put(0,0){\includegraphics[scale=0.9]{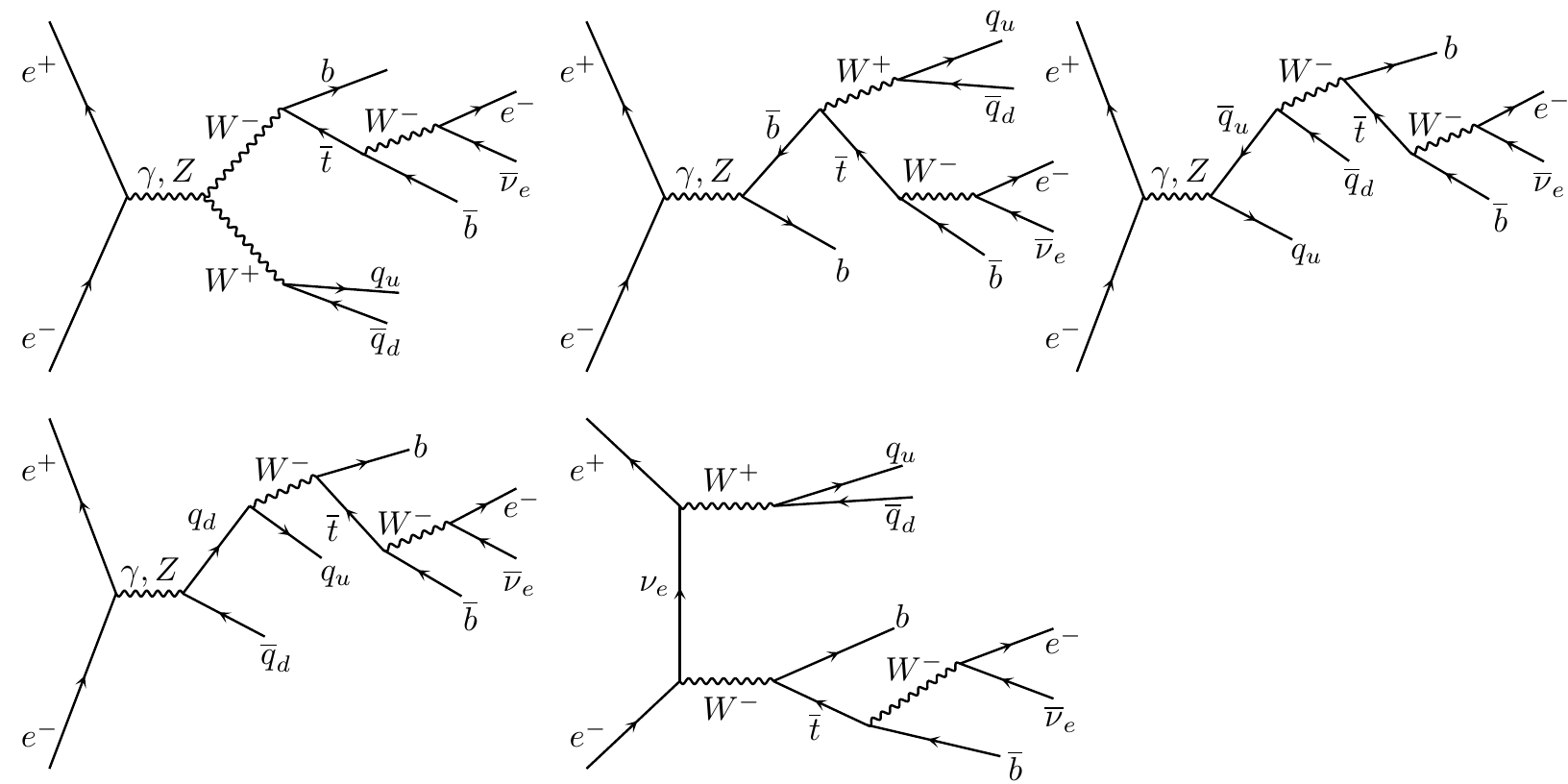}}
\end{picture}
\caption{Unitary gauge Feynman diagrams for single-top production in $e^-e^+$
  collisions with $s$-channel vector boson exchange and leptonic top
  decay.}  
  \label{fig:feyn3}
\end{figure}

\begin{figure}[ht!]
\centering
\begin{picture}(150,100)
\put(0,0){\includegraphics[scale=0.9]{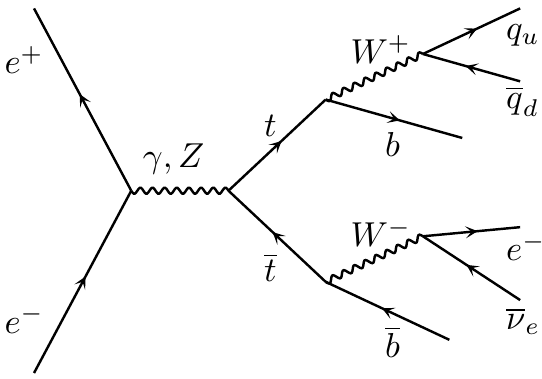}}
\end{picture}
\caption{Unitary gauge Feynman diagrams with a top-pair intermediate state.  These
  diagrams contribute to single-top production when one top line is on
shell and the other off shell.}  
  \label{fig:feyn4}
\end{figure}

The Feynman diagrams for the process (\ref{eq:prddcy}) are shown in
figures \ref{fig:feyn1}--\ref{fig:feyn4} for the final states
containing $e^-\bar{\nu}_e$.  We set the electron mass $m_e=0$, thus
decoupling the electron from the Higgs field. We take into account
only Cabibbo mixing in our computations, since third-generation mixing
can be safely neglected for our purposes. Thus, in (\ref{eq:prddcy})
we have $(q_u,\bar{q}_d)=(u,\bar{d})$, $(u,\bar{s})$, $(c,\bar{d})$,
$(c,\bar{s})$.  With these considerations, the 6 topologies
corresponding to $t$-channel vector boson exchange in figure
\ref{fig:feyn1} lead to 40 Feynman diagrams.  Notice that those
diagrams involve only hadronic top decay.  $s$-channel vector boson
exchange diagrams with one internal top line decaying hadronically are
given by the 5 topologies in figure \ref{fig:feyn2}, corresponding to
32 diagrams. $s$-channel diagrams with one top decaying leptonically
are given by the 5 topologies in figure \ref{fig:feyn3}, leading to 36
diagrams. Finally, figure \ref{fig:feyn4} shows one topology,
corresponding to 8 diagrams for $s$-channel vector boson exchange with
two internal top lines, which contribute to single-top production when
one top line is on its mass shell and the other one off shell.  We
have, then, a total of 116 diagrams for semileptonic single-top
production and decay in the SM with Cabibbo mixing, in the
$e^-\bar{\nu}_e$ channel.  If full CKM mixing is taken into account
the number of diagrams doubles to 232, since the additional diagrams
with third-generation mixing can be obtained from the ones without it
by just exchanging the $\bar{q}_d$ and $\bar{b}$ final-state lines in
each diagram.


\begin{figure}[ht!]
\centering
\begin{picture}(250,290)
  \put(0,0){\includegraphics[scale=1.0]{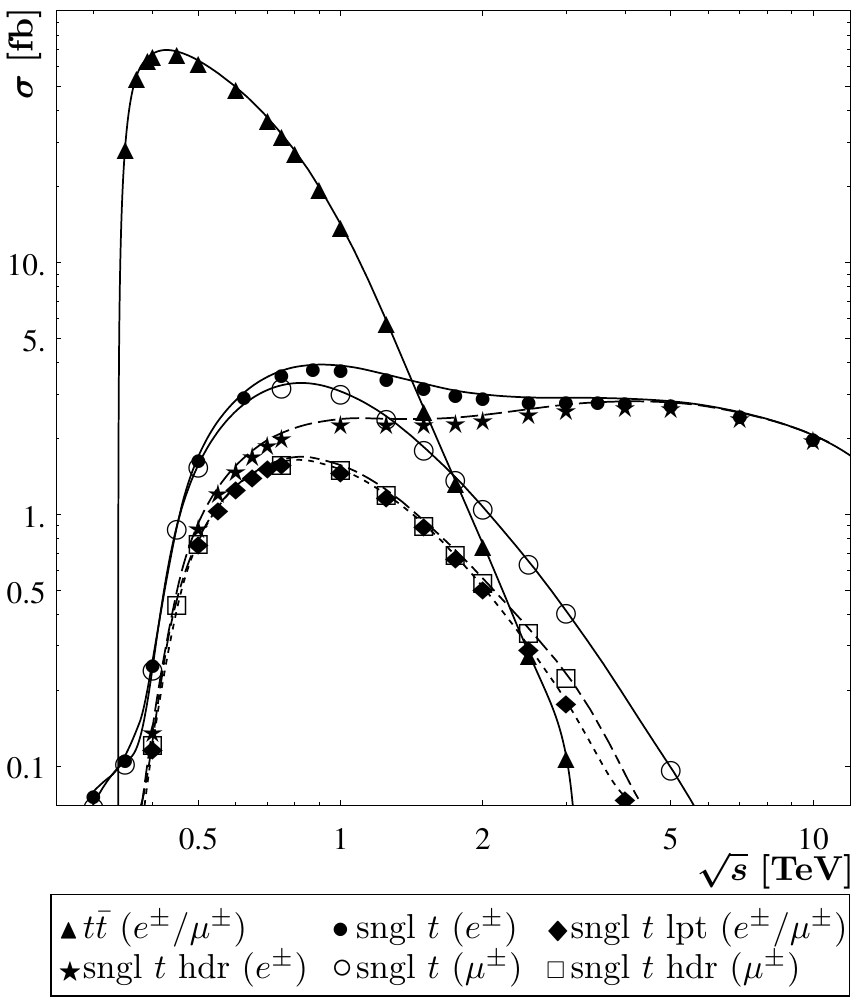}}
\end{picture}
\caption{Total cross section dependence with $\sqrt{s}$ for the
  single-top and top-pair production processes (\ref{eq:phsprgn}) with
minimal phase-space cuts (\ref{eq:a0}). Also shown are processes with
$\mu\nu_\mu$ in the final state, which proceed only through
the $s$-channel diagrams in figures \ref{fig:feyn2}--\ref{fig:feyn4}.}
  \label{fig:sqrts1}
\end{figure}

We consider the processes (\ref{eq:prddcy}), given by the
Feynman diagrams in figures \ref{fig:feyn1}--\ref{fig:feyn4},
restricted to the following phase-space regions,
\begin{equation}
  \label{eq:phsprgn}
  \begin{array}{rl}
  \mbox{single top hadronic:} & m(be\nu)\not\in
  I_t\;\mathrm{and}\;m(bjj)\in I_t,\\
  \mbox{single top leptonic:} & m(be\nu)\in I_t\;\mathrm{and}\;m(bjj)\not\in
  I_t,\\  
    \mbox{single top:} & \left( m(be\nu)\in I_t\;\mathrm{and}\;m(bjj)\not\in
    I_t \rule{0pt}{12pt}\right)\;\mathrm{or}\;\left( m(be\nu)\not\in
    I_t\;\mathrm{and}\;m(bjj)\in I_t \rule{0pt}{12pt}\right), \\
  \mbox{top pair:} & m(be\nu)\in I_t\;\mathrm{and}\;m(bjj)\in I_t,\\
  \mbox{offshell $t$:} & m(be\nu)\not\in I_t\;\mathrm{and}\;m(bjj)\not\in I_t,
  \end{array}
\end{equation}
where $I_t$ is a mass interval around the top mass, $I_t=(132,212)$
GeV, and $m(be\nu)$, $m(bjj)$ refer to the invariant mass of the
three-particle sets in the final state that can originate from a top
decay.  The single-top region in (\ref{eq:phsprgn}) corresponds to the
process we are interested in, or ``signal'' process. The top-pair
region in (\ref{eq:phsprgn}) refers to the production of an on-shell
top pair, whereas the line labeled ``off-shell $t$'' corresponds to
the phase-space region where no on-shell top is produced.  Figure
\ref{fig:sqrts1} displays the dependence on $\sqrt{s}$ of the total
cross sections for the processes (\ref{eq:prddcy}) restricted to the
regions (\ref{eq:phsprgn}), with minimal phase-space cuts (see
(\ref{eq:a0}) below).  Also shown in the figure are the cross sections
for the processes with $\mu^\mp$ final states, whose diagrams are
given by those in figures \ref{fig:feyn2}--\ref{fig:feyn4} with the
replacement $e,\nu_e\rightarrow\mu,\nu_\mu$.  The cross sections for
the muonic final states are equal to those for the process
(\ref{eq:prddcy}) restricted to $s$-channel only, so that they
illustrate the role of the $s$-channel in (\ref{eq:prddcy}).

The irreducible background to single-top production consists of all
processes (\ref{eq:procs}) not proceeding through the on-shell
production of a single top.  We distinguis three contributions to the
irreducible background: (\emph{i}) top-pair production, arising from
the diagrams in figure \ref{fig:feyn4} with both top lines on shell,
(\emph{ii}) off-shell top processes, comprising the diagrams in
figures \ref{fig:feyn1}--\ref{fig:feyn4} with all internal top lines
off shell, and (\emph{iii}) no-top processes originating from all
Feynman diagrams for (\ref{eq:procs}) not containing any top quark
propagator.  It is clear by definition that there can be no
interference between the processes (\emph{i}) and (\emph{ii}), and the
interference between (\emph{i}) and (\emph{iii}) turns out to be
suppressed, as discussed below.  Therefore, it is appropriate to adopt
the convention to refer to the contributions (\emph{ii}) and
(\emph{iii}) together as irreducible background, and to (\emph{i}) as
a separate top-pair production background.

For the process (\ref{eq:procs}), with Cabibbo mixing and for the
$e^-\bar{\nu}_e$ channel, there are 2064 diagrams without internal $t$
lines (1808 with 6 electroweak vertices and 256 with 4 electroweak
vertices and 2 strong ones).  As we discuss in more detail below,
at ILC/CLIC energies, 90\% of the irreducible background cross section
stems from the $WH$ and $WZ$ associated production processes
\begin{subequations}
  \label{eq:asoczh}
\begin{alignat}{6}
  \label{eq:asoch}
  e^- e^+ \rightarrow&
  W^+ H e^- \bar{\nu}_e, &\quad&W^+\rightarrow q_u\bar{q}_d,
  &\quad&H\rightarrow b\bar{b},\\
    \label{eq:asocz}
  e^- e^+ \rightarrow&    
  W^+ Z e^- \bar{\nu}_e, &\quad&W^+\rightarrow q_u\bar{q}_d,
  &\quad&Z\rightarrow b\bar{b},      
\end{alignat}
\end{subequations}
and their charge conjugates.  The process (\ref{eq:asoch}) involves 96
Feynman diagrams and (\ref{eq:asocz}) 320, for a total of 416
diagrams.  In the computation of the irreducible background described
below, however, we take into account the full process (\ref{eq:procs}).

\subsection{Phase-space cuts and event selection}
\label{sec:cuts}

We compute the tree-level cross sections for single-top production and
decay, and for the background processes, with the matrix-element Monte
Carlo program \textsc{MadGraph5\_aMC@NLO} (henceforth MG5) version 2.3
\cite{mg5}.  In all cases we set $m_t=172$ GeV, $m_b=4.7$ GeV,
$m_c=1.27$ GeV, $m_Z=91.19$ GeV, $m_W=79.82$ GeV, $m_h=125$ GeV,
$\alpha(m_Z)=1/132.507$, $G_F=1.1664\times10^{-5}$ GeV$^{-2}$,
$\alpha_S(m_Z)=0.118$. The masses of the lighter quarks, $e$ and $\mu$
are set to vanish, and the Higgs vacuum-expectation value $v=246.22$
GeV. Furthermore, we take into account Cabibbo mixing with
$\theta_c=0.228$. For event analysis we use \textsc{Root} version 5.34
\cite{root}.

\begin{figure}[ht!]
\centering
\begin{picture}(280,270)
  \put(0,0){\includegraphics[scale=0.5]{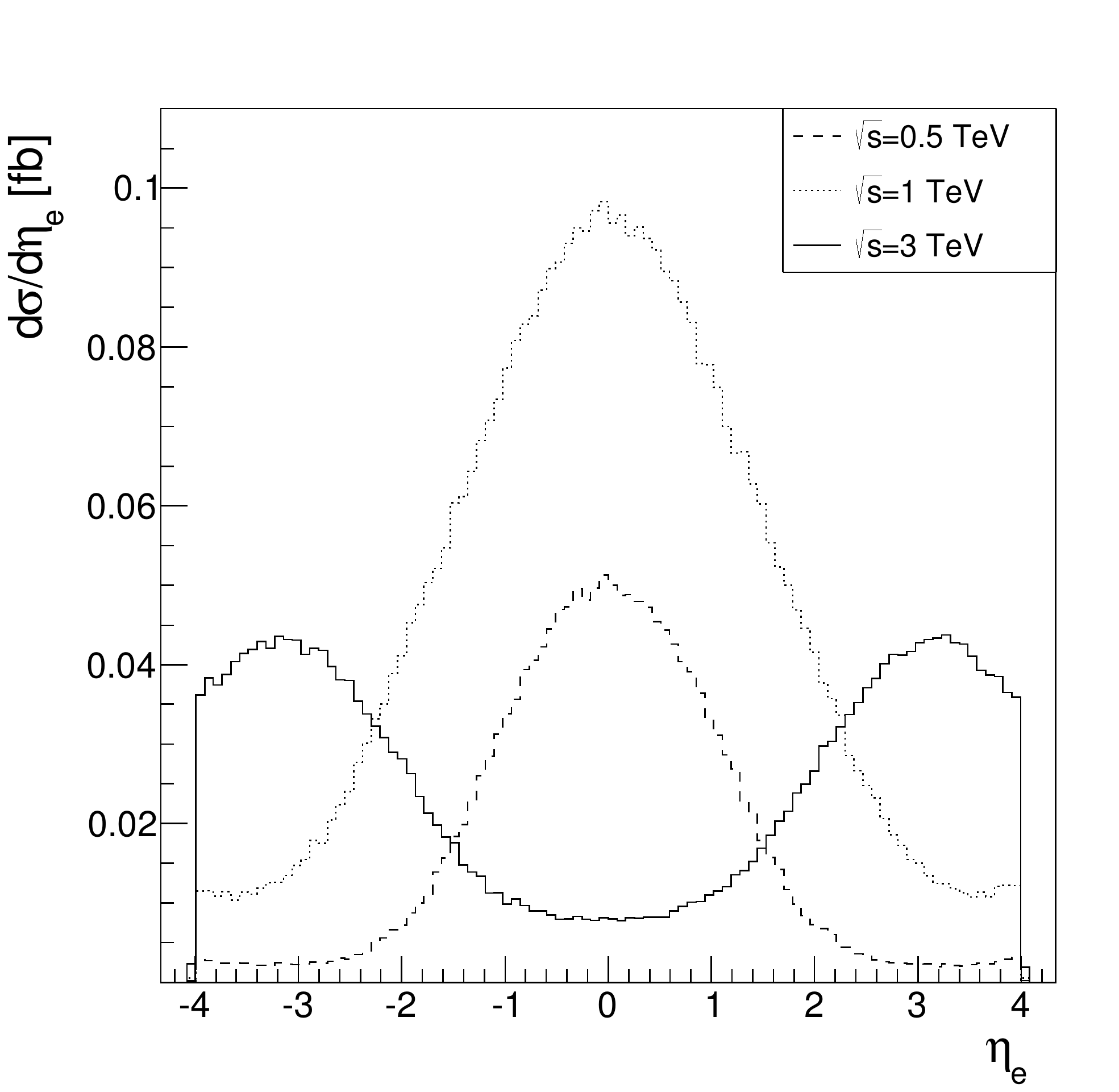}}
\end{picture}
\caption{Differential cross sections for the $e^\mp$ pseudorapidity
  for the signal process (\ref{eq:prddcy}) with the cuts $A_0$,
  (\ref{eq:a0}), and $C_0$, (\ref{eq:c0}), normalized to total cross
  section (see table \ref{tab:tab1}).}
  \label{fig:fgr0}
\end{figure}

In order to make the cross section well defined, and to improve the
signal-to-background ratio, we apply several phase-space cuts
discussed in detail in what follows.  We impose minimal centrality and
isolation cuts in the form
\begin{equation}
  \label{eq:a0}
  A_0:\quad  |\eta(e)|<4, \; |\eta(j)|<3, \; \Delta R_\mathrm{ch.} > 0.5,
\end{equation}
where $\eta(e)$ refers to the pseudorapidity of the final-state
electron or positron, $\eta(j)$ to that of the jets, and $\Delta
R_\mathrm{ch.}$ to the distance in the $\eta$-$\phi$ plane between any
pair of charged particles.  We assume that the central detector system
covers the central region $|\eta|<3$--3.5 and the forward detectors
the region $3<|\eta|<4$, as is expected to be the case at the ILC/CLIC 
\cite{ilc4,clic2}.  As shown in figure \ref{fig:fgr0}, the cut on
$\eta(e)$ is substantially more restrictive at the CLIC energy than at
ILC's.  This is a consequence of the fact that at $\sqrt{s}=3$ TeV
the process (\ref{eq:prddcy}) occurs mostly through the $t$-channel
vector boson exchange diagrams of figure \ref{fig:feyn1}, while at the
ILC the $s$-channel diagrams of figures
\ref{fig:feyn2}--\ref{fig:feyn4} dominate, as shown in figure
\ref{fig:sqrts1}.  The cut on $\Delta
R=\sqrt{\Delta\eta^2+\Delta\varphi^2}$ in (\ref{eq:a0}) is an
isolation cut setting the minimal distance between any two charged
particles in the final state.

\begin{figure}[ht!]
\centering
\begin{picture}(450,450)
\put(-60,-165){\includegraphics[scale=0.95]{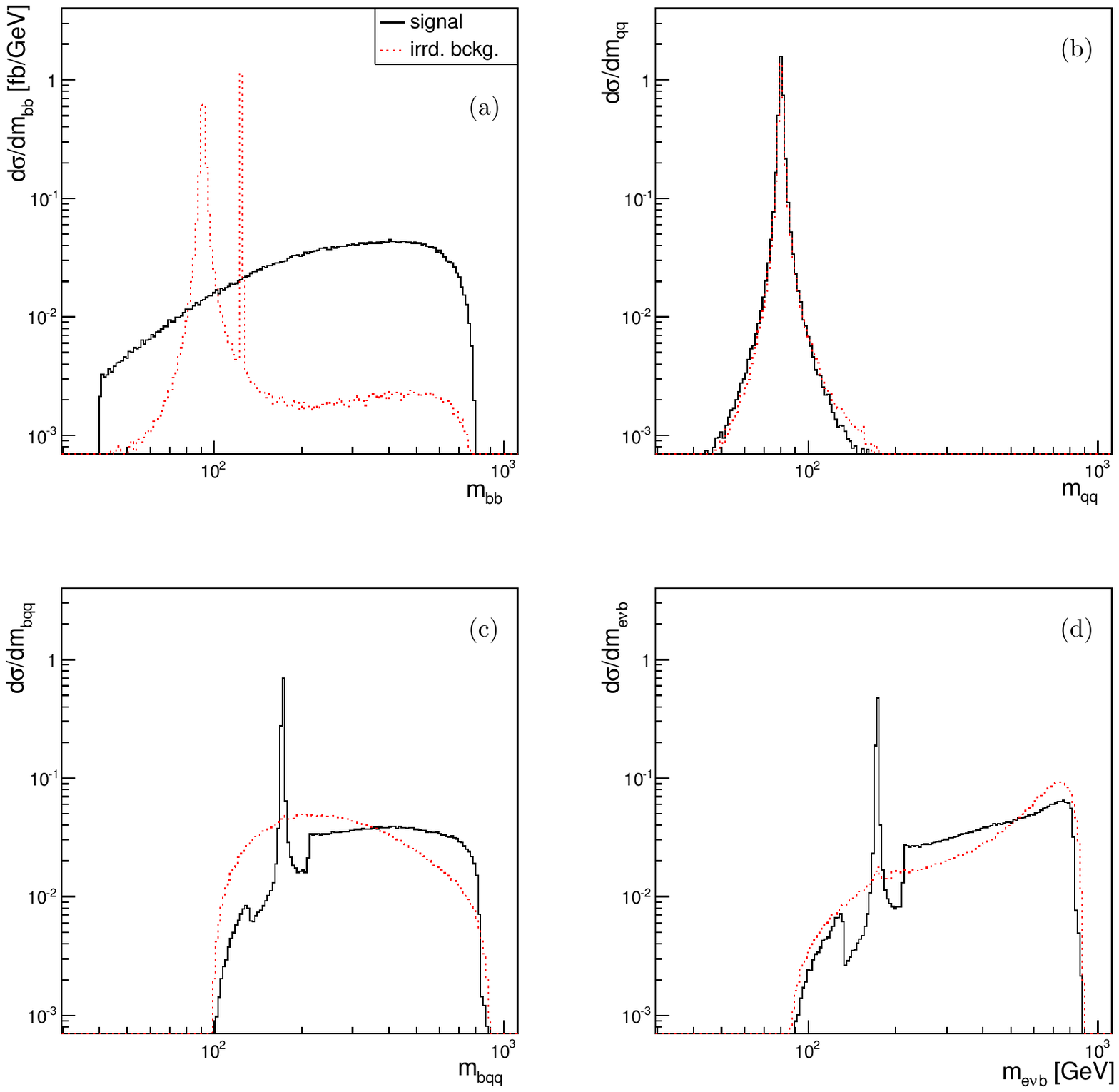}}
\end{picture}
\caption{Differential cross sections for the signal process
  (\ref{eq:prddcy}) and irreducible background at $\sqrt{s}=1$ TeV
  and with the cuts $A_0$, (\ref{eq:a0}),  and $C_0$,
  (\ref{eq:c0}). All differential cross sections are normalized to
  total cross section: 3.58 fb for the signal and 3.13 fb for the
  background processes.}  
  \label{fig:fgr2}
\end{figure}

The reducible background to the process (\ref{eq:prddcy}), which is
studied in more detail below in section \ref{sec:redcbl}, consists of
final states with four jets, an electron/positron and $\not\!\!
E_\perp$, with the number of $b$-jets $N_b\neq2$.  Such final states
contain pairs of massless partons, light quarks or gluons, which lead
to infrared singularities.  In order to avoid those singularities we
require the final-state jets to satisfy the condition
\begin{equation}
  \label{eq:c0}
  C_0:\quad m(j,j')>
  \left\{
      \begin{array}{cccl}
    60\,\mathrm{GeV} &\quad&\mathrm{if}&\sqrt{s}=\mbox{3 TeV},\\
    40\,\mathrm{GeV} &\quad&\mathrm{if}&\sqrt{s}=\mbox{1 TeV},\\
    30\,\mathrm{GeV} &\quad&\mathrm{if}&\sqrt{s}=\mbox{0.5 TeV},
      \end{array}
\right.      
\end{equation}
where $m(j,j')$ refers to the mass of any pair of partons in the final
state.  As shown in table \ref{tab:tab1}, the restriction $C_0$
cuts about 35\% of the irreducible background cross sections both at 
ILC and CLIC energies, and it cuts about 10\% of the signal at the ILC
and 30\% at the highest CLIC energy. 
\newsavebox{\dash}
\savebox{\dash}{\mbox{---}}
\newcommand{\udsh}{\usebox{\dash}}
\begin{table}[ht!]
  \centering
  \begin{tabular}{c|ccc|c|c}\hline
    \multicolumn{5}{c}{$\sigma$ [fb], $\sqrt{s}=0.5$ TeV}      \\\hline
          &sngl-h&sngl-l&sngl&pair & irr.bkg.\\\hline
    $A_0$ &0.93  &0.79  &1.72&62.28&2.09     \\\hline
    $C_0$ &0.85  &0.72  &1.58&58.07&1.78     \\\hline
    $C_1$ &0.53  &0.46  &0.99&35.41&1.10     \\\hline
    $C_2$ &0.53  &0.46  &0.99&\udsh&0.47     \\\hline
    $C_3$ &0.36  &0.32  &0.68&\udsh&0.039    \\\hline
    $C_4$ &0.34  &0.29  &0.63&\udsh&0.026    \\\hline    
    \multicolumn{5}{c}{$\sqrt{s}=1$ TeV}      \\\hline
    $A_0$ &2.42  &1.53  &3.93&14.09&4.50     \\\hline
    $C_0$ &2.15  &1.43  &3.58&12.94&3.13     \\\hline
    $C_1$ &1.31  &0.89  &2.29&7.87 &1.94     \\\hline
    $C_2$ &1.31  &0.89  &2.29&\udsh&1.07     \\\hline
    $C_3$ &1.06  &0.78  &1.85&\udsh&0.074    \\\hline
    $C_4$ &0.99  &0.72  &1.72&\udsh&0.057    \\\hline        
    \multicolumn{5}{c}{$\sqrt{s}=3$ TeV}      \\\hline
    $A_0$ &2.77  &0.18  &2.95&0.12 &11.25    \\\hline
    $C_0$ &1.90  &0.16  &2.06&0.096&7.15     \\\hline
    $C_1$ &1.16  &0.099 &1.26&0.059&4.43     \\\hline
    $C_2$ &1.16  &0.099 &1.26&\udsh&1.40     \\\hline
    $C_3$ &0.96  &0.096 &1.06&\udsh&0.060    \\\hline
    $C_4$ &0.93  &0.087 &1.02&\udsh&0.045    \\\hline    
  \end{tabular}                               
  \caption{Effect of the  phase-space cuts
    (\ref{eq:a0})--(\ref{eq:c4}) on the total cross section for the
    process (\ref{eq:prddcy}) and its 
    subprocesses (\ref{eq:phsprgn}). The $b$-tagging efficiency and
    mistagging probabilities involved in $C_1$ are given in the text
    under equation (\ref{eq:c1}).}
  \label{tab:tab1}
\end{table}
In figure \ref{fig:fgr2} we display several mass differential cross
sections computed with the cuts (\ref{eq:a0}), (\ref{eq:c0}), for the
signal and irreducible background at $\sqrt{s}=1$ TeV. At the other
energies, 0.5 and 3 TeV, the distributions are qualitatively
similar. The distributions of the $b$-pair mass $m(b\bar{b})$ and the
light-jet pair $m(q\bar{q})$ for the irreducible background (dotted
lines in figures \ref{fig:fgr2} (a) and (b), resp.)
are seen to be dominated by the $Z$, $h$ peaks and $W$ peak,
respectively, as expected from the main irreducible-background
processes (\ref{eq:asoczh}).  For instance, at $\sqrt{s}=3$ TeV the
total cross section for (\ref{eq:asoczh}) with the cuts $A_0$ and
$C_0$ is found to be 6.57 fb, amounting to 92\% of that of the total
irreducible backgroung, 7.15 fb, as given in table \ref{tab:tab1}.

To suppress the reducible background we require the final state to
contain exactly two $b$-tagged jets:
\begin{equation}
  \label{eq:c1}
  C_1:\quad N_b=2. 
\end{equation}
We assume the $b$-jet tagging efficiency to be $\eta_b=80\%$, the
mistagging probabilities to be $p_c=10\%$ for $c$-jets and $p_q=1\%$
for the lighter jets.  These are realistic values, consistent with the
efficiencies currently achieved by LHC detectors \cite{cms12}.  In our
analysis we simulate $b$ tagging by relabeling $b$ quarks as lighter
ones with 20\% probability and by relabeling as $b$ quarks $c$ quarks
with 10\% probability and lighter partons with $1\%$ probability.
Thus, cut (\ref{eq:c1}) results in the signal and irreducible
background cross sections being reduced to about 61\% of their value.
The cut $C_1$ plays an important role in rejecting the reducible
background, as discussed in detail in the following section.

Most of the rejection of the irreducible background, and of the
reducible background events remaining after cut (\ref{eq:c1}), is
achieved by requiring that each event must contain the decay products
of exactly one on-shell top quark, as described by the ``single top''
line in (\ref{eq:phsprgn}).  We denote the four jets in the final
state by $J_{0,\ldots,3}$, with $J_{0,1}$ the two $b$-tagged jets and
$J_{2,3}$ the two light jets, the latter ones in no particular order.
In events with a hadronically decaying top, we denote the promptly
produced $b$ jet as $J_0$ and the $b$ jet from top decay as $J_1$, and
for leptonic top decays we denote the prompt $b$ jet as $J_1$ and the
$b$ jet from top decay as $J_0$.  Thus, with this convention and with
the notation introduced in (\ref{eq:phsprgn}), the phase space cut for
single-top events can be written as
\begin{equation}
  \label{eq:c2}
 C_2:\quad \left( m(J_0,e,\nu)\in I_t\;\mathrm{and}\;m(J_1,J_2,J_3)\not\in
    I_t \rule{0pt}{12pt}\right)\;\mathrm{or}\;\left( m(J_0,e,\nu)\not\in
    I_t\;\mathrm{and}\;m(J_1,J_2,J_3)\in I_t \rule{0pt}{12pt}\right), 
\end{equation}
where the first parenthesis corresponds to leptonically decaying and
the second to hadronically decaying top quarks. In figure
\ref{fig:fgr2} (c), the differential cross section for the invariant
mass $m(b,j,j)$ for the signal process (solid line) corresponds to
$m(J_1,J_2,J_3)$, and in \ref{fig:fgr2} (d),
$m(b,e,\nu)$ corresponds to $m(J_0,e,\nu)$.  For the irreducible
background, the distributions of $m(b,j,j)$ and $m(b,e,\nu)$ obtained
with each one of the two $b$ quarks in the final state are essentially
identical.

The irreducible background contains a substantial combinatorial
component satisfying (\ref{eq:c2}), as is apparent from figures
\ref{fig:fgr2} (c) where the distribution of
$m(b,q,q)$ is seen to contain a significant number of events under the
top mass peak.  This yields a sizable irreducible background cross
section even after the cut $C_2$ has been applied, as shown in table
\ref{tab:tab1}.  Since most of the irreducible
background consists of processes (\ref{eq:asoczh}) in which the
$b\overline{b}$ pair comes from $Z$ or $h$ decay, we are led to
introduce the cut 
\begin{equation}
  \label{eq:c3}
  C_3:\quad \left\{
      \begin{array}{cccl}
    m(J_0,J_1)>130\,\mathrm{GeV} &\quad&\mathrm{if}&\sqrt{s}\geq\mbox{1 TeV},\\
    m(J_0,J_1)>130\quad\textrm{or}\quad m(J_0,J_1)<70 \,\mathrm{GeV}
    &\quad&\mathrm{if}&\sqrt{s}=\mbox{0.5 TeV}, 
      \end{array}
\right.      
\end{equation}
to further suppress the remaining irreducible background. Finally,
we require that the two light jets be the decay products of an
on-shell $W$ boson (cf.\ figures \ref{fig:feyn1}--\ref{fig:feyn4}) and
that the final state contains substantial $\not\!\!E_T$,
\begin{equation}
  \label{eq:c4}
  C_4:\quad 60\,\mathrm{GeV} < m(J_2,J_3) < 100\,\mathrm{GeV},
  \quad \mbox{and} \quad
  \not\!\!E_T > 10\,\mbox{GeV}.
\end{equation}
Cut $C_4$ provides some further suppression of the irreducible
background without significantly affecting the signal.

The cumulative effect of the phase-space cuts
(\ref{eq:a0})--(\ref{eq:c4}) on signal and background is shown in
table \ref{tab:tab1}.  The lines of this table labeled $C_4$ give the
cross sections for the signal and irreducible background, including
the effect of $b$-tagging efficiencies and all phase-space cuts.  From
those results, and assuming an integrated luminosity $L=1$ ab$^{-1}$,
we estimate the statistical uncertainty in the signal cross section
$\sigma_\mathrm{sgn}$ to be 4.0\%, 2.4\%, 3.1\% at $\sqrt{s}=0.5$, 1,
3 TeV, respectively. The irreducible background is seen to amount to
4.1\%, 3.3\%, 4.4\% of $\sigma_\mathrm{sgn}$, respectively, at the
same three energies. 

\subsection{Top-pair background}
\label{sec:ttxbck}

Table \ref{tab:tab1} shows that the top-pair production background is
fully rejected by the cut $C_2$, but that is true only when
measurement uncertainties are not taken into account.  If we allow for
the effect on the cut $C_2$ of measurement uncertainties in
$m(J_1,J_2,J_3)$ and $m(J_0,e,\nu)$, however, a fraction of top-pair
events would pass that cut. For reasonably small measurement
uncertainties we expect that fraction of events to be a relatively
small fraction of the pair-production cross section.  In the case of
CLIC, in which the $t\bar{t}$ cross section is already very small
compared to that of single-top, this effect is expected to
be of second-order.  At the ILC, however, top-pair production
is substantially larger than single-top, so that even a small
fraction of these events can become a large background.
In this section we quantify the top-pair background to
single-top production taking into account the top-mass reconstruction
uncertainty.

A goal of the ILC and CLIC detectors is to achieve high enough jet
energy- and mass-resolution to be able to separate the $W$ and $Z$
peaks in the dijet mass spectrum \cite{ilc4,clic2}.  As shown in
figure 2.6 of \cite{clic2}, for that purpose the mass resolution
$\sigma_m/m$ must be better than 5\%, with good separation of the $W$
and $Z$ peaks obtained for $\sigma_m/m=2.5\%$.  Motivated by that
observation, we assume that the top mass will be reconstructed from
three jets at the ILC/CLIC with a relative uncertainty in the range
5--7.5\%.  In order to obtain a quantitative estimate of the effect of
measurement uncertainties on the $t\bar{t}$ background, we randomly
smear the reconstructed masses $m(J_1,J_2,J_3)$ and $m(J_0,e,\nu)$
before applying the cut $C_2$ to each top-pair production event.  We
assume those masses to be independently normal-distributed with
standard-deviation parameter $\sigma_m=\veps_m m$, where $\veps_m$ is
the assumed relative uncertainty.  For simplicity we assume the same
value of $\veps_m$ for both $m(J_1,J_2,J_3)$ and $m(J_0,e,\nu)$.  We
carry out this randomized analysis of the $t\bar{t}$ event sample a
few thousand times to obtain a statistical sample of the $t\bar{t}$
cross section after cuts $A_0$, $C_{0,\ldots,4}$.  The resulting
$\sigma_{t\bar{t}}$ distribution is strongly asymmetric with a long
tail to the right, so we characterize it by the interval
$(\langle\sigma_{t\bar{t}}\rangle, \langle\sigma_{t\bar{t}}\rangle +
\Delta\sigma_{t\bar{t}})$ .

At $\sqrt{s}=3$ TeV, if we assume the top mass to be reconstructed
with uncertainty $\veps_m=5\%$, the $t\bar{t}$ background turns out to
be 0.48\%--0.50\% of the single-top cross section with all cuts,
$A_0$, $C_{0,\ldots,4}$, as given in table \ref{tab:tab1}.  For
$\veps_m=7.5\%$, we find the $t\bar{t}$ background to be
0.51\%--0.79\% of the single-top cross section.  As expected, that
background turns out to represent a small uncertainty at CLIC energy.

At $\sqrt{s}=1$ TeV, for $\veps_m=5\%$, the $t\bar{t}$ background
result is 1.6\%--3.3\% of the single-top cross section.  For
$\veps_m=7.5\%$, we get 3.5\%--16.4\%.  We see that for $\veps_m$ less
than 7.5\% the $t\bar{t}$ background is limited to about 15\% of the
single-top cross section, and for lower values near $\veps_m\simeq
5\%$ that background can be somewhat less than 5\%.

At $\sqrt{s}=0.5$ TeV, for $\veps_m=5\%$, the $t\bar{t}$ background is
8.6\%--17\% of the single-top cross section.  If $\veps_m=7.5\%$, we
get 16.5\%--50.5\% for the $t\bar{t}$ background. These results are
also not unexpected, since the $t\bar{t}$ cross section is large at
the ILC at 0.5 TeV and can potentially swamp the single-top process.
They also suggest, however, that the top-pair background can be
limited to the range 10--30\% for $\veps_m$ less than 7\%, and to the
range 10--20\% if $\veps_m$ is close to 5\%.

\subsection{Reducible background}
\label{sec:redcbl}

The reducible background to the single-top production process
(\ref{eq:prddcy}) is given by processes of the form
\begin{equation}
  \label{eq:redbck}
    e^-e^+ \rightarrow 
 j_1j_2j_3j_4\,e^-\overline{\nu}_e 
+
 j_1j_2j_3j_4\,e^+\nu_e,
\end{equation}
with $j=u,d,c,s,b,g$ or their antiparticles. The number of
$b$/$\bar{b}$ jets in (\ref{eq:redbck}) can be $N_b=0,1,3$ (with
$N_b=2$ corresponding to the signal and irreducible background
processes (\ref{eq:procs}), and $N_b=4$ being forbidden by electric
charge conservation).  Since the probability to mistag a final state
(\ref{eq:redbck}) with $N_b\neq2$ as one with $N_b=2$ as in
(\ref{eq:procs}) depends on $N_b$ and on the number $N_c$ of
$c$/$\bar{c}$ quarks, we have to consider separately the cases with
different values of $N_b,N_c$. 

For our computation of the reducible background we adopt
two-generation Cabibbo mixing, since the effects of third-generation
mixing on cross sections are numerically inconsequential.  This
implies in particular that only processes (\ref{eq:redbck}) with
$N_b=0$ and without internal $t$ lines are possible.  Indeed, explicit
computation shows that the final states (\ref{eq:redbck}) with
$N_b=1$, 3, which can only occur through third-generation mixing, lead
to cross sections of $\mathcal{O}(10^{-2}\mathrm{fb})$ at most even
before the restrictive cuts $C_{1\ldots4}$ in equations
(\ref{eq:c1})--(\ref{eq:c4}) are applied.  For the same reason,
equally negligible cross sections are obtained for diagrams with
$N_b=0$ with one or more internal $t$ lines.

\begin{figure}[ht!]
\centering
\begin{picture}(450,400)
\put(0,0){\includegraphics[scale=0.85]{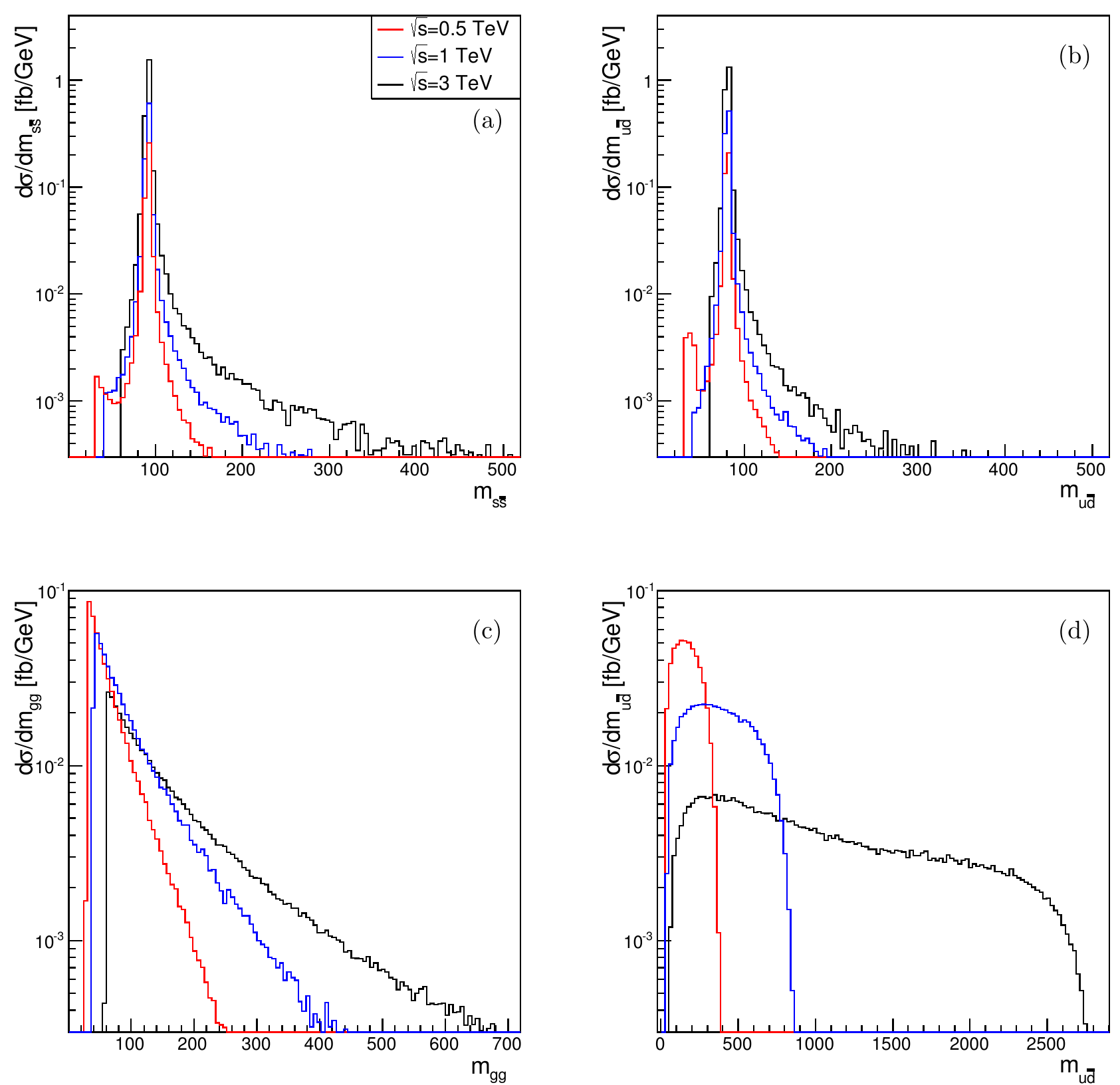}}
\end{picture}
\caption{Differential cross sections for the reducible background
  processes $e^-e^+\rightarrow u\bar{d}s\bar{s} e^-\bar{\nu}_e$ ((a)
  and (b)) and  $e^-e^+\rightarrow u\bar{d}gg e^-\bar{\nu}_e$ ((c) and
  (d)), with the cuts $A_0$, (\ref{eq:a0}),  and $C_0$,
  (\ref{eq:c0}), normalized to total cross section.}  
  \label{fig:fgr4}
\end{figure}

The reducible bakground (\ref{eq:redbck}) contains a large number of
subprocesses whose detailed description is not needed for our
purposes. With only two-generation mixing taken into account, it
involves 15632 Feynman diagrams with $e^-\bar{\nu}_e$ final states.
However, some general features of this background, with cuts $A_0$,
$C_0$, are easily understood. With those cuts the gluon final states
$q_u\bar{q}_dgg$ are a minority fraction of the cross section. The
final states $q_u\bar{q}_dq\bar{q}$ which dominate the cross section
originate mostly from $WZ$ associated production, as seen from figure
\ref{fig:fgr4}, in a similar way as the irreducible background.

We compute the reducible--background cross section by applying
the acceptance cuts $A_0$, (\ref{eq:a0}), to particles with
electric or color charge (i.e., to charged leptons, quarks and
gluons).  The cut $C_0$, (\ref{eq:c0}), is also required to make the
cross section infrared finite.  Simulation of light-parton mistagging
is carried out, as discussed in the text below (\ref{eq:c1}), by
relabeling $c$ quarks as $b$ ones with probability $p_c=0.1$ and
lighter partons with probability $p=0.01$.  The resulting mistagging
probabilities for processes (\ref{eq:redbck}) with $N_b=0$ are
given by
\newcommand{\ten}[1]{$\times10^{-#1}$}
\begin{equation}
  \label{eq:mistag}
    \begin{array}{ccrl}
      0_b0_c:&\quad& 6p^2q^2                    &=5.88\times10^{-4},\\
      0_b1_c:&\quad& 3pp_cq^2+3p^2qq_c          &=3.21\times10^{-3},\\
      0_b2_c:&\quad& p_c^2q^2+4pp_cqq_c+p^2q_c^2&=1.34\times10^{-2},\\
      0_b3_c:&\quad& 3p_c^2qq_c+3pp_cq_c^2      &=2.92\times10^{-2},
    \end{array}
\end{equation}
with $q=1-p$, $q_c=1-p_c$.  The mistagged final states, containing two
``fake'' $b$ quarks, are then required to pass the cuts
$C_{2\ldots4}$, as defined in (\ref{eq:c2})--(\ref{eq:c4}), in the
same way as the signal and irreducible background.  The effects of
cuts $A_0$, $C_{0,1}$ on the total cross sections is shown in table
\ref{tab:tab5}.  Notice that, at each energy, the cross sections
corresponding to cut $C_1$ can be obtained by multiplying the results
for the cuts $A_0$ and $C_0$ by the corresponding probabilities in
(\ref{eq:mistag}), up to a small numerical uncertainty.  As seen from
the table, those cross sections are at most
$\mathcal{O}(10^{-2}\mathrm{fb})$.  After cuts $C_{2\ldots4}$ are
applied, the resulting reducible-background cross sections are at most
$\mathcal{O}(10^{-3}\mathrm{fb})$ and therefore negligible.
\begin{table}[ht!]
  \centering
  \begin{tabular}{c|cccc}\hline
    \multicolumn{5}{c}{$\sigma$ [fb], $\sqrt{s}=0.5$ TeV}      \\\hline
          &$0_b0_c$&$0_b1_c$&$0_b2_c$&$0_b3_c$      \\\hline\hline
                $A_0,C_0$ &1.71       &1.72       &0.38       &0.38   \\\hline
\rule{0pt}{11pt}    $C_1$ &1.03\ten{3}&5.43\ten{3}&5.14\ten{3}&1.11\ten{2}\\\hline
    \multicolumn{5}{c}{$\sqrt{s}=1$ TeV}      \\\hline
          &$0_b0_c$&$0_b1_c$&$0_b2_c$&$0_b3_c$      \\\hline\hline
                $A_0,C_0$ &3.29       &3.29       &0.85       &0.85   \\\hline
\rule{0pt}{11pt}    $C_1$ &2.04\ten{3}&9.48\ten{3}&1.12\ten{2}&2.42\ten{2}\\\hline
    \multicolumn{5}{c}{$\sqrt{s}=3$ TeV}      \\\hline
          &$0_b0_c$&$0_b1_c$&$0_b2_c$&$0_b3_c$      \\\hline\hline
                $A_0,C_0$ &6.85       &6.82       &2.00       &1.98       \\\hline
\rule{0pt}{11pt}    $C_1$ &3.50\ten{3}&1.98\ten{2}&2.49\ten{2}&5.35\ten{2}\\\hline
\end{tabular}                               
  \caption{Effect of the  phase-space cut
    (\ref{eq:c1}) on the total cross section for the
    process (\ref{eq:redbck}). The $b$-tagging efficiency and 
    mistagging probabilities involved in $C_1$ are given in the text
    under equation (\ref{eq:c1}).}
  \label{tab:tab5}
\end{table}

\subsection{Beam polarization}
\label{sec:polar}

The ILC baseline design supposes a polarization of the electron beam
of at least 80\%, and 30\% for the positron beam
\cite{ilc3b}. 
For CLIC, the baseline design assumes an electron beam polarization of
80\% and an unpolarized positron beam \cite{clic2}. 
In both accelerators a later upgrade is foreseen that would increase
the positron beam polarization to 60\% \cite{ilc3b,clic2}.  In the
study of single-top production beam polarization may lead to a
reduction of the measurement uncertainties, either by increasing the
signal cross section, therefore reducing the statistical uncertainty,
or by suppressing important backgrounds.

At $\sqrt{s}=3$ TeV the cross-section uncertainties are dominated by
the irreducible background and the statistical uncertainty (about
4.4\% and 3.1\%, respectively, as noted at the end of section
\ref{sec:cuts}). The top-pair production background uncertainty is
subdominant, as discussed in section \ref{sec:ttxbck}.  The
longitudinal beam polarization combinations that cause the signal
cross section to increase or decrease have the same effect on the
irreducible background, though not necessarily in the same amount.
With longitudinal polarizations $(\P_{e^-},\P_{e^+})=(-90\%,60\%)$
both the signal and irreducible background increase, leading to a
statistical uncertainty of 2\% and an irreducible background of 6.2\%,
thus worsening the overall uncertainty.  For
$(\P_{e^-},\P_{e^+})=(-90\%,-60\%)$ we get a statistical uncertainty
of 3.2\% and an irreducible background of 4.7\%, and for
$(\P_{e^-},\P_{e^+})=(90\%,60\%)$ we get 4.4\% and 4.5\%,
respectively; these cases show little change in the uncertainties with
respect to the unpolarized case.  Finally, for
$(\P_{e^-},\P_{e^+})=(90\%,-60\%)$ the statistical uncertainty grows
to 8\% while the irreducible background decreases to 3.3\%, for an
overall uncertainty considerably worse than the unpolarized result.
At this energy, therefore, we find that even the highest possible beam
polarizations do not lead to a significant reduction in the cross
section uncertainty with respect to the unpolarized case.

At $\sqrt{s}=1$ TeV, assuming an integrated luminosity $L=1$
ab$^{-1}$, we found in section \ref{sec:cuts} a statistical
uncertainty of 2.4\% and an irreducible background of 3.3\%.  As
discussed in section \ref{sec:ttxbck}, top-pair production
may be significantly larger than the irreducible background.
Assuming the highest possible beam polarizations, we find that
$(\P_{e^-},\P_{e^+})=(-90\%,60\%)$ increases the signal cross section
$\sigma_\mathrm{sgnl}$ by a factor 2.72 relative to the unpolarized
case, thus reducing the statistical uncertainty to 1.5\%.  The
irreducible and top-pair production backgrounds increase by a factor
of 2 and 2.24, respectively, so that
$\sigma_\mathrm{irr}/\sigma_\mathrm{sgnl}$ decreases by 25\% and
$\sigma_{t\bar{t}}/\sigma_\mathrm{sgnl}$ by 20\% with respect to the
unpolarized case.  The polarizations
$(\P_{e^-},\P_{e^+})=(-90\%,-60\%)$ also reduce the ratio
$\sigma_{t\bar{t}}/\sigma_\mathrm{sgnl}$, but by decreasing all cross
sections. The signal cross section $\sigma_\mathrm{sgnl}$ decreases
by a factor 0.79, thus slightly increasing the statistical uncertainty
to 2.7\%.  The ratio $\sigma_\mathrm{irr}/\sigma_\mathrm{sgnl}$
remains unchanged, and $\sigma_{t\bar{t}}/\sigma_\mathrm{sgnl}$
decreases by 25\% with respect to the unpolarized case.

At $\sqrt{s}=0.5$ TeV from section \ref{sec:cuts}
we get a statistical uncertainty of 4\% and an irreducible background
of 3.3\% relative to the signal cross section $\sigma_\mathrm{sgnl}$.
From section \ref{sec:ttxbck} it is apparent that at this energy the
$t\bar{t}$ production background strongly dominates the cross section
uncertainty.  The beam polarizations
$(\P_{e^-},\P_{e^+})=(-90\%,60\%)$ increases $\sigma_\mathrm{sgnl}$ by
a factor of 2.7 and $\sigma_{t\bar{t}}$ by 2.2, thus leading to a
decrease in $\sigma_{t\bar{t}}/\sigma_\mathrm{sgnl}$ of 20\%.  The
polarizations $(\P_{e^-},\P_{e^+})=(-90\%,-60\%)$ decrease
$\sigma_\mathrm{sgnl}$ by a factor 0.7 and $\sigma_{t\bar{t}}$ by 0.6,
thus reducing $\sigma_{t\bar{t}}/\sigma_\mathrm{sgnl}$ by 20\%.

Therefore, at ILC energies we expect the highest beam polarizations to
lead to a moderate reduction of single-top cross section uncertainties
by 20--25\%.  Lower polarizations would yield correspondingly smaller
uncertainty reductions.


\section{Effective operators for single-top production at
  $\boldsymbol{e^-e^+}$ colliders}
\label{sec:singletoperators}

The Lagrangian for the single-top production processes (\ref{eq:prddcy})
is of the form
\begin{equation}
  \label{eq:lag}
  \mathcal{L} = \mathcal{L}_\mathrm{SM} + \frac{1}{\Lambda^2}\sum_{\mathcal{O}}
  (C_\mathcal{O} \mathcal{O}+\mathrm{h.c.})+\cdots, 
\end{equation}
where $\mathcal{O}$ denotes dimension 6 effective operators, $\Lambda$
is the new-physics scale, and the ellipsis refers to
higher-dimensional operators. It will be convenient in what follows to
express our results in terms of the modified dimensionless couplings
\begin{equation}
  \label{eq:coupl}
  \bar{C}_\mathcal{O} = C_\mathcal{O} \frac{v^2}{\Lambda^2},
\end{equation}
where $v$ is the Higgs-field vacuum expectation value.  At tree level
the coupling constants $\C_\mathcal{O}$ are independent of the scale
$\Lambda$. We denote complex couplings as
$\C_{\mathcal{O}}=\C_{\mathcal{O}\,r}+i \C_{\mathcal{O}\,i}$.

Throughout this paper we use the dimension 6 effective operators from
the operator basis given in \cite{grz10}.  However, we adopt the sign
convention in the covariant derivatives and the operator normalization
defined in \cite{zha14}, where a factor $y_t$ is attached to an
operator for each Higgs field it contains, and a factor $g$ ($g'$) for
each $W_{\mu\nu}$ ($B_{\mu\nu}$) field-strength tensor.  We are
interested in those operators that can contribute to single top
production at an $e^+e^-$ collider.  There are operators with flavor
changing couplings, but in this study we will not consider them. 

\subsection{Operators that generate $\boldsymbol{tbW}$ couplings.}
\label{sec:wtbops}

There are four dimension 6 operators in the basis \cite{grz10} that
give rise to effective $tbW$ couplings: $O_{\varphi q}^{(3)}$,
$O_{uW}$, $O_{\varphi ud}$ and $O_{dW}$, where we are omitting
generation indices.  The first two also generate neutral current (NC)
couplings and are among the operators contributing to $e^+e^- \to t\bar
t$.  We combine operator $O^{(3)}_{\vp q}$ with
$O^{(1)}_{\vp q}$ so as to eliminate the $bbZ$ neutral current term.
Expanding these operators in physical fields we obtain: 
\bea O_{\vp q}^{(-)33} &=& O_{\vp q}^{(3)33}-O_{\vp q}^{(1)33} =
\frac{y_t^2}{\sqrt{2}}g (v+h)^2 \left( W^+_\mu \bar t_{L} \gm b_L +
  W^-_\mu \bar b_{L} \gm t_L \right)
\nonumber \\
&+& y_t^2\frac{g}{c_W} (v+h)^2 Z_\mu \bar t_{L}\gm t_{L} \, ,
\nonumber \\
O^{33}_{\vp ud} &=& \frac{y_t^2}{2\sqd}g (v+h)^2 W^+_\mu \bar t_R \gm
b_{R} \,,
\nonumber \\
O^{33}_{uW} &=& 2 y_t g (v+h) \left( \partial_\mu W^-_\nu + ig W^3_\mu
  W^-_\nu \right) \bar b_{L} \smn t_R
\nonumber \\
&+& \sqd y_t g (v+h) \left( \partial_\mu W^3_\nu + ig W^-_\mu W^+_\nu
\right) \bar t_{L} \smn t_R \, ,
\nonumber \\
O^{33}_{dW} &=& 2 y_t g (v+h) \left( \partial_\mu W^+_\nu + ig W^+_\mu
  W^3_\nu \right) \bar t_L \smn b_{R}
\nonumber \\
&-& \sqd y_t g (v+h) \left( \partial_\mu W^3_\nu + ig W^+_\mu W^-_\nu
\right) \bar b_{L} \smn b_{R} \, .
  \label{eq:operators}
\eea

The relation between the couplings $\bar C_{\mathcal{O}}$
in (\ref{eq:coupl}) and the usual $\delta
  V_L$, $V_R$, $g_L$ and $g_R$ $tbW$ form factors is: 
\begin{equation}
  \label{wtbfactors}
V_L = V_{tb} + y^2_t \bar C^{(-)33}_{\vp q},\quad
V_R = y^2_t \frac{1}{2} \bar C^{33}_{\vp ud},\quad
g_L = -y_t g \sqd \bar C^{33}_{dW},\quad
g_R = -y_t g \sqd \bar C^{33}_{uW}.
\end{equation}
There have been many studies that have set bounds on the coefficients
of these operators based mostly on LHC ($\sqrt s=7,$ 8 TeV)
single top production and $W$-helicity fractions in top decay
\cite{escobar,romero,cms17,globalfits,englert16,cms2017,atlas2017}.
A recent LHC combined extraction of $|f_{LV}V_{tb}|$ has been
presented at 1$\sigma$ level \cite{escobar}: 
\bea 
|f_{LV}V_{tb}| = 1.02 \; \pm 0.08 ({\rm meas}) \; \pm 0.04 ({\rm
  theo}) \;,  
\eea 
where $f_{LV} V_{tb}\equiv V_L$ in our notation.  From this we can set
$-0.16 < \bar C_{\vp q}^{(-)33} < 0.20$ at 2$\sigma$ (or 95\%) level,
if we assume $V_{tb}=1$.  Concerning the other couplings, with the
constraint $V_L=1$, CMS has reported a global analysis based on two
and three-dimensional fit scenarios, from which they have obtained the
most stringent $95\%$ C.L.\ bounds to date \cite{cms17}:
\begin{equation}
  \label{eq:cms17limits}
|V_R| < 0.16,\quad 
|g_L| < 0.057,\quad 
-0.049 < g_R < 0.048.  
\end{equation}
This limits can be converted to bounds on the effective couplings
$\C_\mathcal{O}$ through (\ref{wtbfactors}), 
\begin{equation}
  \label{eq:cmslimfin}
-0.16<\C^{(-)33}_{\vp q}<0.20,\quad
|\C^{33}_{\vp ud}| < 0.32,\quad
-0.053<\C^{33}_{uW}<0.052,\quad
|\C^{33}_{dW}|<0.062,
\end{equation}
which are therefore the current LHC bounds on the effective
dimension-6 couplings $\C_{\mathcal{O}}$.

What could be the improvement of these limits in the LHC top physics
program, assuming no BSM physics is found?  With integrated
luminosities of up to 20 fb$^{-1}$ already obtained, the statistical
uncertainties are subdominant relative to the systematic and other
uncertainties.  We could then expect that even the HL-LHC phase will
not necessarily yield an order-of-magnitude improvement over the
limits (\ref{eq:cms17limits}), (\ref{eq:cmslimfin}).  For instance, in
\cite{englert17} the $95\%$C.L.\ limits $|\bar C_{\vp q}^{(-)33}|<0.5$
and $|\bar C_{uW}^{33}|<0.25$ are obtained based on CMS and ATLAS
$ttZ$ cross section measurements.  A projection is then made in
\cite{englert17} for as much as $3{\rm ab}^{-1}$ of pseudodata leading
to the estimates $|\bar C_{\vp q}^{(-)33}|<0.2$ and $|\bar
C_{uW}^{33}|<0.15$, that amount to an improvement by a factor of 2.

The ILC $t\bar t$ production process has the potential of improving
the $\bar C_{\vp q}^{(-)33}$ and $\bar C_{uW}^{33}$ bounds by an order
of magnitude or more \cite{agsa12,englert17,amjad}.  Specifically, as
reported in \cite{agsa12}, the ILC at $\sqrt s = 500$ GeV based on
$t\bar t$ cross section and forward-backward (FB) asymmetry
measurements, and assuming an experimental uncertainty of 5\% in the
cross section and 2\% in the FB asymmetry, would give the
single-coupling bounds,
\bea
|\bar C_{\vp q}^{(-)33}| < 0.04,\quad
|\bar C_{uW}^{33}| < 0.006.
\label{aguilarlimits}
\eea
In \cite{englert17}, for the same energy an experimental uncertainty
of 1\% is assumed, which yields correspondingly tighter 
individual-coupling direct bounds,
\bea
|\bar C_{\vp q}^{(-)33}| < 0.015,\quad
|\bar C_{uW}^{33}| < 0.0011,
\label{englertlimits}
\eea 
as shown in figure 6 of that reference. Notice that
(\ref{englertlimits}) assumes also longitudinal beam polarizations
$(\P_{e^-},\P_{e^+})=(-80\%,+30\%)$ \cite{englert17}. To bear in mind,
there are other two operators with $ttZ$ couplings that also
contribute to $t\bar t$ production.  When the simultaneous
contributions of these operators are considered, the marginalised
limits are relaxed.  This is indeed so with the marginalised limits
from the multivariate analysis carried out in \cite{englert17}, which
are reported there to be larger than the individual ones
(\ref{englertlimits}) by a factor 17.  For the purposes of the present
study, however, we take the single-coupling bounds
(\ref{englertlimits}) and (\ref{aguilarlimits}) as benchmarks of the
projected sensitivity of $t\bar{t}$ production at the ILC.

\subsection{Diagonal four fermion operators}
\label{sec:diag4}

The use of effective Lagrangians in the top quark physics program aims
ultimately to constrain simultaneusly all the non-redundant operators
(at a certain level, like dimension 6) based on all the available
experimental measurements.  For instance, a recent global fit has been
presented in \cite{englert16} where 4 top-gauge boson and 5
four-fermion operator coefficients were constrained using both $t\bar
t$ as well as single top production measurements from the LHC and the
Tevatron.  Not surprisingly, the bounds obtained by considering effects
from one operator at a time tend to greatly relax when other operators
are also taken into account \cite{englert16}.  There is indeed a great
effort to perform global fit studies as is found in the literature
\cite{globalfits}.  Besides the motivation for making an analysis
complete, the goal of considering all the operators is for
consistency.  Top-gauge boson operators with derivatives on fermion
fields \cite{agsa09b} do not appear in the basis of
\cite{grz10} because equations of motion relate them to the ones
considered here.  These equations involve four-fermion terms that are part
of the non-redundant operators.  For instance, a $tbW$ coupling is
generated by an operator $O_{qW}=\bar q_L \gamma^\mu \tau^I D^\nu q_L
W^I_{\mu \nu}$ that should be considered in a CC interaction of the
top quark.  Bearing in mind the general $tbW$ vertex generated by the
operators in (\ref{eq:operators}), we can in fact isolate the
nonredundant contribution by this operator \cite{bach12}.  However, it
is convenient to implement this effect with the four-fermion operator
$O^{(3)}_{lq}$ that generates a $\bar e_L \gamma^\mu \nu_L \bar t_L
\gamma_\mu b_L$ interaction instead \cite{bach12}:
\begin{equation}
  \label{oqwrelation}
O_{qW}^{33} + ( O_{qW}^{33} )^\dagger =  \frac{g}{2}
\left(   O_{\vp q}^{(3)33} + ( O_{\vp q}^{(3)33} )^\dagger
\right)
+ \frac{g}{2} \sum^3_{k=1} \left( {O}^{(3)kk33}_{\ell q} +
{O}^{(3)kk33}_{qq} \right).
\end{equation}
In this sense, an analysis of top-gauge boson couplings should be
considered complete and consistent only if it includes four-fermion
operators. 

There are 8 diagonal four-fermion operators involving two
first-generation leptons and two third-generation quarks.  Four of
them are associated to CC couplings, and the other four to only NC
couplings.  The CC operators that contribute to single top production
are: 
\bea O^{(3)13}_{\ell q} &=& \bar \ell \gamma_\mu \tau^I \ell \bar
q \gm \tau^I q \;\; = \; 2 \left( \bar \nu_L \gamma_\mu e_L \bar b_L
  \gm t_L \, +\, \bar e_L \gamma_\mu \nu_L \bar t_L \gm b_L \right)
\nonumber \\
&+& \left( \bar \nu_L \gamma_\mu \nu_L \, - \, \bar e_L \gamma_\mu e_L
\right) \; \left( \bar t_L \gamma_\mu t_L \, -\, \bar b_L \gamma_\mu
  b_L \right)
\nonumber \\
O^{13}_{\ell edq} &=& \bar \nu_L e_R \bar b_R t_L \, +\, \bar e_L e_R
\bar b_R b_L
\nonumber \\
O^{(1)13}_{\ell equ} &=& \bar \nu_L e_R \bar b_L t_R \, -\, \bar e_L
e_R \bar t_L t_R
\nonumber \\
O^{(3)13}_{\ell equ} &=& \bar \nu_L \smn e_R \bar b_L \sigma_{\mu \nu}
t_R \, -\, \bar e_L \smn e_R \bar t_L \sigma_{\mu \nu} t_R,
\label{eq:operators4q}
\eea
where we have used the abbreviated notation $O^{13}$ for $O^{1133}$.

To date, there are no reported limits based on LHC nor
Fermilab top production and/or decay processes for these
operators.  In \cite{zha14} it has been pointed out
that distribution-based measurements like the $W$-helicity
fractions could in principle be used to this end as the
leptonic decay is used to analyse the $W$-polarization.
However, the experimental requirement that $m_{\ell \nu}$
be close to $M_W$ severely reduces the sensitivity of
the fractions to the four fermion operators \cite{zha14}.


\section{Effective couplings in single-top production and decay}
\label{sec:res}

For the computation of the cross section with anomalous effective
vertices we use MG5, as described in section \ref{sec:cuts}. For
computational purposes we set the scale $\Lambda=$ 10 TeV in
(\ref{eq:lag}), (\ref{eq:coupl}). The effective operators
(\ref{eq:operators}), (\ref{eq:operators4q}) were implemented in MG5
by means of the program \textsc{FeynRules} version 2.0 \cite{feynrul}.

\begin{figure}[ht]
  \centering
\includegraphics[scale=0.825]{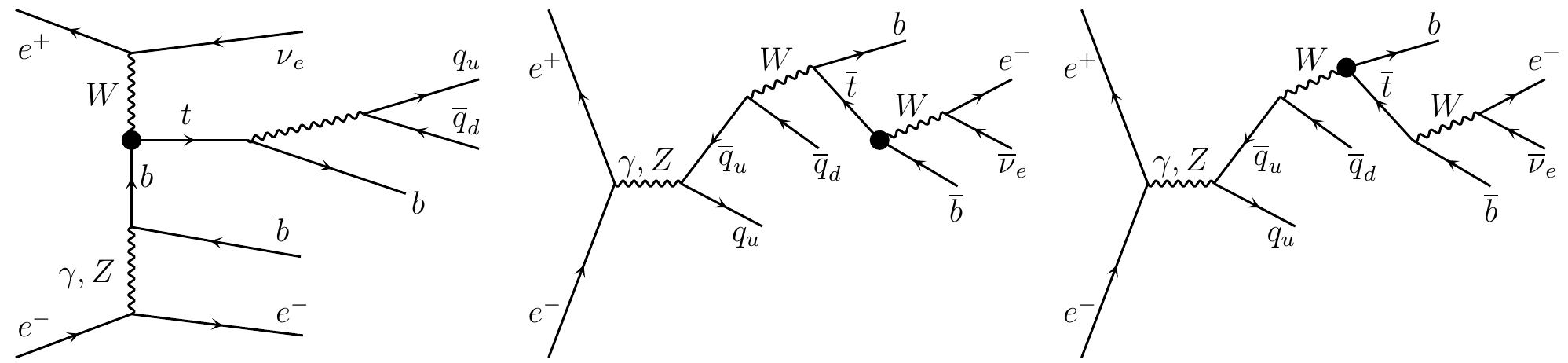}  
  \caption{Sample Feynman diagrams with one anomalous three-particle
    vertex from the operators (\ref{eq:operators}). } 
  \label{fig:fgra1}
\end{figure}

\begin{figure}[ht]
  \centering
\includegraphics[scale=0.875]{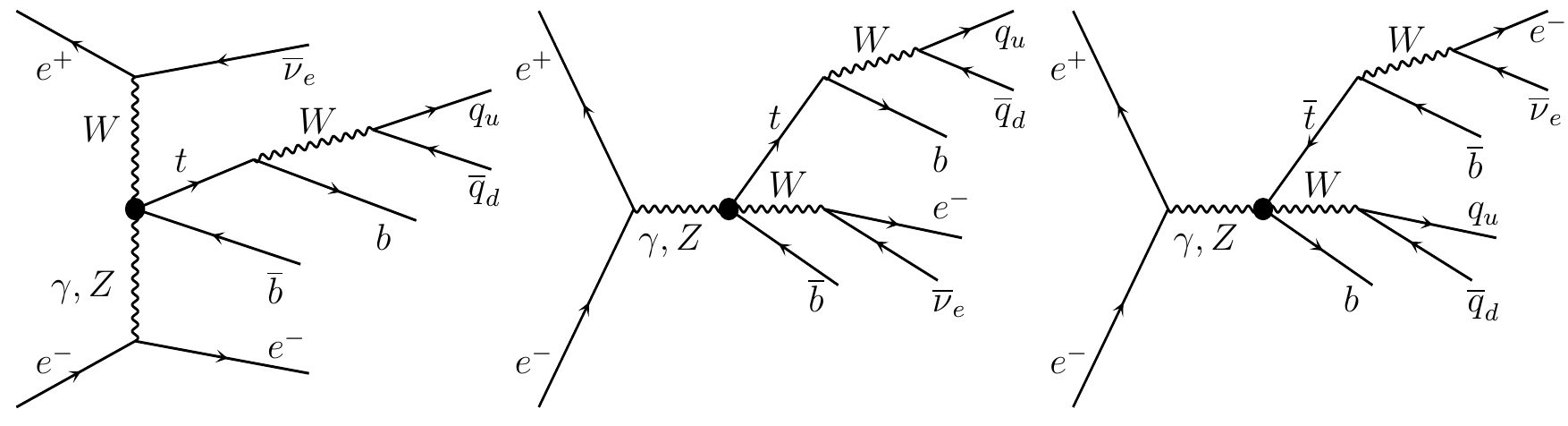}  
\caption{All Feynman diagrams with one anomalous four-particle vertex
  from the operators $O_{uW}^{33}$, $O_{dW}^{33}$ in (\ref{eq:operators}).}
  \label{fig:fgra2}
\end{figure}

\begin{figure}[ht]
  \centering
\includegraphics[scale=0.8,trim=0 160 0 0,clip=true]{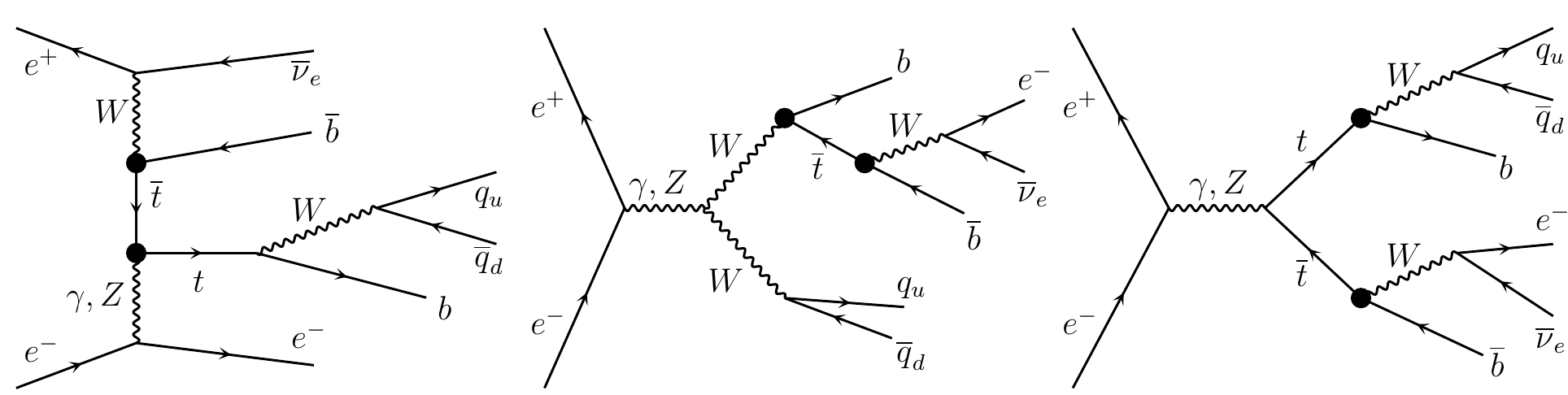}  
\caption{Sample Feynman diagrams with two and three anomalous vertices
  from the operators (\ref{eq:operators}).}
  \label{fig:fgra3}
\end{figure}

\begin{figure}[ht]
  \centering
\includegraphics[scale=0.92]{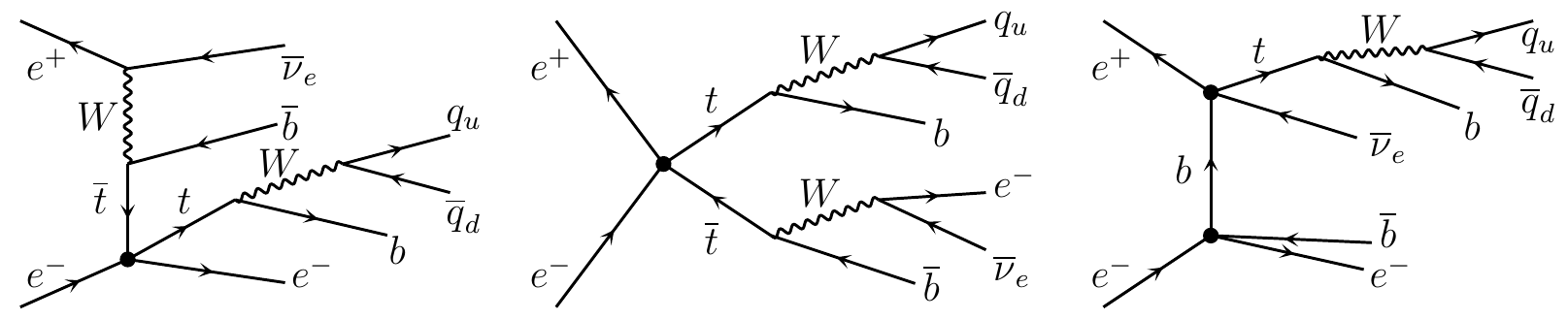}  
\caption{Sample Feynman diagrams with one and two four-fermion
  vertices from the operators (\ref{eq:operators4q}).}
  \label{fig:fgr4f}
\end{figure}

\begin{figure}[ht]
  \centering
\includegraphics[scale=0.92]{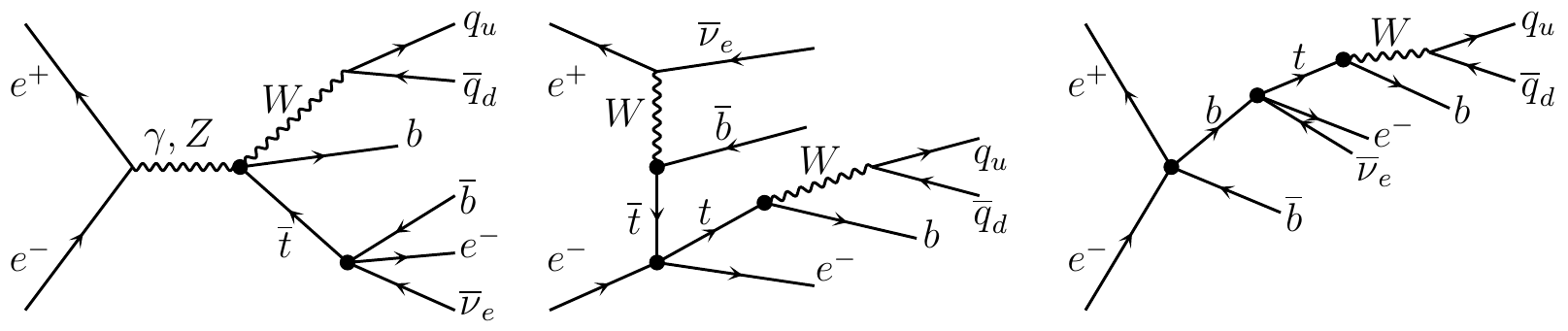}  
\caption{Sample Feynman diagrams with both gauge-boson and
  four-fermion vertices from the operators (\ref{eq:operators}),
  (\ref{eq:operators4q}).} 
  \label{fig:fgr4f2}
\end{figure}

The Feynman diagrams for the processes (\ref{eq:prddcy}) containing
effective vertices are illustrated in figure
\ref{fig:fgra1}--\ref{fig:fgr4f2}.  As discussed in section
\ref{sec:sgnlirr}, there are 116 diagrams for semileptonic single-top
production and decay in the SM with Cabibbo mixing, in the
$e^-\bar{\nu}_e$ channel.  When the effective operators are switched
on in Lagrangian (\ref{eq:lag}), there are 296 additional diagrams
with one vertex from the operators (\ref{eq:operators}) and none from
the operators (\ref{eq:operators4q}), or $N_{tbW}=1$, $N_{4f}=0$; 220
with $N_{tbW}=2$, $N_{4f}=0$; 40 with $N_{tbW}=3$, $N_{4f}=0$; 124
with $N_{tbW}=0$, $N_{4f}=1$; 24 with $N_{tbW}=0$, $N_{4f}=2$; 192
with $N_{tbW}=1$, $N_{4f}=1$; 20 with $N_{tbW}=1$, $N_{4f}=2$; and 60
with $N_{tbW}=2$, $N_{4f}=1$, for a total of 976 additional diagrams.

Diagrams with one, two and three effective vertices entering the
amplitude for (\ref{eq:prddcy}), contribute to it at
$\mathcal{O}(\Lambda^{-n})$ with $n=2,$ 4 and 6, respectively. In
fact, once the top propagator dependence on effective couplings
through the top decay width is taken into account, the scattering
amplitude is given as a power series of $\Lambda^{-2}$.  We remark
that diagrams with two effective vertices must be kept in the
amplitude since, through their interference with SM diagrams, they
make contributions to the cross section of the same order,
$\mathcal{O}(\Lambda^{-4})$, as the square of diagrams with only one
effective vertex.  We have actually taken into account the
contributions from diagrams with three effective vertices in our
calculation as well as the dependence of the top decay width on the
effective couplings, but we have explicitly verified in all cases that
the contribution to the cross section from terms of order higher than
$\mathcal{O}(\Lambda^{-4})$ is actually negligible for values of the
effective couplings within the bounds given below.  (We remark here,
parenthetically, that the contributions to the cross section at order
$1/\Lambda^4$ from dimension 8 operators interfering with the SM are
currently unknown and constitute an inherent uncertainty of the EFT
analysis at dimension 6.)

\subsection{Methodology and assumptions}
\label{sec:method}

In order to obtain bounds on the effective couplings, we consider the
ratio of the cross section $\sigma_\mathrm{eff}(\{\C_\mathcal{O}\})$
obtained from the Lagrangian (\ref{eq:lag}) at tree level to the SM
cross section $\sigma_\mathrm{SM}=\sigma_\mathrm{eff}(\{0\})$ 
\begin{equation}
  \label{eq:rat}
  R=\frac{\sigma_\mathrm{eff}(\{\C_\mathcal{O}\})}{\sigma_\mathrm{SM}},
\end{equation}
where $\{\C_\mathcal{O}\}$ is the set of anomalous coupling constants.
For a given relative experimental uncertainty $\veps_\mathrm{exp}$,
the region of allowed values for the effective couplings
$\{\C_\mathcal{O}\}$ is determined at the 1$\sigma$ level by the
inequalities
\begin{equation}
  \label{eq:ineq}
  R \lessgtr 1 \pm \veps_\mathrm{exp}.
\end{equation}
We obtain allowed intervals on the effective couplings taken to be
non-zero one at a time by parameterizing the ratio (\ref{eq:rat}) as
\begin{equation}
  \label{eq:rat1}
  R = 
  1+a \C_\mathcal{O} + b \C_\mathcal{O}^2 + \cdots,
\end{equation}
where the ellipsis refers to higher powers of $\C_\mathcal{O}$.
Similarly, we consider also allowed two-coupling regions for pairs
of effective couplings by parameterizing (\ref{eq:rat}) as
\begin{equation}
  \label{eq:rat2}
  R= 
  1+a \C_\mathcal{O} + b \C_\mathcal{O}^2 +a' \C_\mathcal{O'} + b'
  \C_\mathcal{O'}^2 + c \C_\mathcal{O}\C_\mathcal{O'} + \cdots,
\end{equation}
with $\C_\mathcal{O}$ and $\C_\mathcal{O'}$ the effective couplings
under consideration, and all other ones set to zero.  The parameters
in (\ref{eq:rat1}) and (\ref{eq:rat2}) are determined from an extensive
set of MG5 simulations to which (\ref{eq:rat1}) and (\ref{eq:rat2}) are
fitted.  Once those parameters are known, (\ref{eq:ineq}) yields the
desired one- or two-dimensional limits on the effective couplings
being considered.  The consistency condition that the contribution to
the cross section from terms of $\mathcal{O}\left(\Lambda^{-6}\right)$
and higher in (\ref{eq:lag}) be negligibly small entails on the
parameterizations (\ref{eq:rat1}), (\ref{eq:rat2}) the requirement that
the terms of $\mathcal{O}(\C^{3})$ and higher must be
correspondingly negligible within the allowed region determined by
(\ref{eq:ineq}).  We check this consistency condition in all cases
considered below.

In order to obtain bounds on the effective couplings through
(\ref{eq:ineq}), below we assume $\veps_\mathrm{exp}$ to take values
within certain intervals.  We motivate those assumed ranges for
$\veps_\mathrm{exp}$ by estimating the uncertainties in the signal
cross section in our Monte Carlo simulations, through the addition in
quadrature of the statistical uncertainty and the irreducible
background given at the end of section \ref{sec:cuts}, and the
uncertainties arising from the $t\bar{t}$ background as discussed in
section \ref{sec:ttxbck}. This leads us to assume that at $\sqrt{s}=3$
TeV $\veps_\mathrm{exp}$ is in the range 5\%--9\%, and at $\sqrt{s}=1$
TeV within 5\%--15\%.  In the case of $\sqrt{s}=0.5$ TeV we assume
$\veps_\mathrm{exp}$ to lie in the interval 10\%--20\%, but present
also some results at 30\%.  As discussed in more detail below, at all
three energies the individual-coupling limits we obtain can be
extrapolated to values of $\veps_\mathrm{exp}$ moderately lower or
higher than the ranges we assume.  Furthermore, as discussed in
section \ref{sec:polar}, beam polarization may contribute to reduce
background uncertainties, which may help reach the lower end of the
assumed uncertainty intervals, especially at $\sqrt{s}=0.5$ TeV.

\subsection{Results}
\label{sec:results}

The single-coupling bounds obtained from (\ref{eq:ineq}), are reported
in table \ref{tab:sngl.cplng} for effective $tbW$ couplings and table
\ref{tab:sngl.cplng.4q} for effective four-fermion couplings.  The
validity of the quadratic dependence (\ref{eq:rat1}) of the cross
section on the effective couplings is verified to hold for all
couplings at the three energies and three experimental uncertainties
shown in the tables, with one exception. The lower bound for $\C_{\vp
  q}^{(-)33}$ at $\sqrt{s}=3$ TeV at the highest value of
$\varepsilon_\mathrm{exp}$ lies outside the interval of validity of
the quadratic approximation (\ref{eq:rat1}), which we estimate to be
$-0.12<\C_{\vp q}^{(-)33}<0.12$ at that energy. That value is
therefore omitted from table \ref{tab:sngl.cplng}.  There is, however,
no loss of relevant information in that omission, since we see from
table \ref{tab:sngl.cplng} that the cross section at ILC energies is
more sensitive than at CLIC to $\C_{\vp q}^{(-)33}$.

\renewcommand{\tabcolsep}{0pt}
\begin{table}[th]
  \centering
  \begin{tabular}{c|ccc|ccc|ccc}
$\sqrt{s}$ [TeV] & \multicolumn{3}{c}{0.5} & \multicolumn{3}{|c}{1} & \multicolumn{3}{|c}{3} \\\hline 
$\veps_\mathrm{exp}$ (\%) & 10 & 15 & 20 & 5 & 10 & 15 & 5 & 7 & 9 \\\hline\hline
$\C_{\vp q}^{(-)33}$    \e{-0.033}{0.056}\e{-0.057}{0.076}\e{-0.083}{0.095}\e{-0.031}{0.015}\e{-0.056}{0.036}\e{-0.082}{0.057}\e{-0.030}{0.054}\e{-0.067}{0.066}\e{\mbox{--- }}{0.077}\\\hline
$\C_{\vp ud\,r}^{33}$\e{-0.90}{0.85}\e{-1.05}{1.00}\e{-1.19}{1.14}\e{-0.32}{0.31}\e{-0.50}{0.49}\e{-0.63}{0.62}\e{-0.21}{0.20}\e{-0.24}{0.23}\e{-0.26}{0.25}\\\hline
\rule{0pt}{14pt}\!
$\C_{\vp ud\,i}^{33}$&$\pm0.87$&$\pm1.03$&$\pm1.16$&$\pm0.31$&$\pm0.50$&$\pm0.63$&$\pm0.21$&$\pm0.24$&$\pm0.26$\\\hline
$\C_{uW\,r}^{33}$\e{-0.013}{0.021}\e{-0.022}{0.028}\e{-0.032}{0.035}\e{-0.013}{0.0058}\e{-0.025}{0.014}\e{-0.041}{0.021}\e{-0.046}{0.020}\e{-0.050}{0.023}\e{-0.053}{0.027}\\\hline
\rule{0pt}{14pt}\!
$\C_{uW\,i}^{33}$&$\pm0.12$&$\pm0.14$&$\pm0.15$&$\pm0.030$&$\pm0.050$&$\pm0.060$&$\pm0.030$&$\pm0.034$&$\pm0.038$\\\hline
$\C_{dW\,r}^{33}$          \e{-0.16}{0.13}\e{-0.19}{0.16}\e{-0.21}{0.18}\e{-0.033}{0.028}\e{-0.050}{0.046}\e{-0.063}{0.058}\e{-0.025}{0.025}\e{-0.029}{0.029}\e{-0.032}{0.032}\\\hline
\rule{0pt}{14pt}\!
$\C_{dW\,i}^{33}$&$\pm0.15$&$\pm0.17$&$\pm0.19$&$\pm0.031$&$\pm0.048$&$\pm0.061$&$\pm0.025$&$\pm0.029$&$\pm0.032$\\\hline
  \end{tabular}
  \caption{Single-coupling limits on effective $tbW$ couplings, for
    three values of $\sqrt{s}$ and three assumed experimental
    uncertainties.} 
  \label{tab:sngl.cplng}
\end{table}

When the interference of diagrams containing one
effective vertex from the operator $\mathcal{O}$ with those from the
SM vanishes, or is suppressed by a small mass parameter, the linear
term in (\ref{eq:rat}) is suppressed and the bounds on
$\C_{\mathcal{O}}$ are symmetric about the origin.  This is the case,
in particular, for the couplings $\C_{\mathcal{O}\,i}$ associated with
the antihermitian part of $\mathcal{O}$, $i/2
(\mathcal{O}-\mathcal{O}^\dagger)$, since those operators are $CP$-odd
and cannot interfere with the SM operators which are
$CP$-even. (Notice that we are only taking into account Cabibbo mixing
in this paper, so that all of the parameters in the SM charged current
are real.)  On the other hand, when the interference of diagrams with
a vertex from the hermitian part of $\mathcal{O}$ with the SM is
suppressed, the bounds on $\C_{\mathcal{O}\,r}$ and
$\C_{\mathcal{O}\,i}$ are the same, and are denoted
$\C_{\mathcal{O}\,r|i}$ in table \ref{tab:sngl.cplng} and
$\F_{\mathcal{O}\,r|i}$ in table \ref{tab:sngl.cplng.4q}. 

At $\sqrt{s}=0.5$, for an experimental uncertainty
$\veps_\mathrm{exp}=30\%$ we obtain the individual-coupling limits
\begin{equation}
  \label{eq:bnd30}  
\begin{gathered}
-0.14<\C_{\vp q}^{(-)33}<0.13,\quad  
-1.42<\C_{\vp ud\,r|i}^{33}<1.37,\quad  
\left|\C_{\vp ud\,i}^{33}\right|<1.40,\\  
-0.055<\C_{uW\,r}^{33}<0.049,\quad
\left|\C_{uW\,i}^{33}\right|<0.19,\quad
-0.25<\C_{dW\,r}^{33}<0.22,\quad
\left|\C_{dW\,i}^{33}\right|<0.23.\quad
\end{gathered}
\end{equation}
At that energy the signal cross section is most sensitive to $\C_{\vp
  q}^{(-)33}$ and $\C_{uW\,r}^{33}$, for which the bounds in
(\ref{eq:bnd30}) are close to the current limits set by CMS as quoted
above in (\ref{eq:cmslimfin}).  Thus, in order to improve on the
current CMS bounds on those couplings, the single-top cross section
should be measured by the ILC at $\sqrt{s}=0.5$ TeV with an
uncertainty $\veps_\mathrm{exp}<30\%$.  We point out also that, at
that energy and with an unrealistically low uncertainty
$\veps_\mathrm{exp}=5\%$, we would obtain the limits $-0.0093<\C_{\vp
  q}^{(-)33}<0.035$.  Comparison with the bounds at the same
uncertainty in table \ref{tab:sngl.cplng}, leads to the conclusion
that the sensitivity to $\C_{\vp q}^{(-)33}$ for a fixed relative
uncertainty is smaller at CLIC than at the ILC.

\begin{table}[th]
  \centering
  \begin{tabular}{c|ccc|ccc|ccc}
$\sqrt{s}$ [TeV] & \multicolumn{3}{c}{0.5} & \multicolumn{3}{|c}{1} & \multicolumn{3}{|c}{3} \\\hline 
$\veps_\mathrm{exp}$ (\%) & 10 & 15 & 20 & 5 & 10 & 15 & 5 & 7 & 9 \\\hline\hline
$\F_{\ell q}^{(3)13}\times10^2$   \e{-0.28}{0.48}\e{-0.48}{0.65}\e{-0.69}{0.83}\e{-0.067}{0.033}\e{-0.12}{0.080}\e{-0.18}{0.13}\e{-0.025}{0.045}\e{-0.047}{0.055}\e{-0.082}{0.065}\\\hline
$\F_{ledq\,r|i}^{13}\times10^2$&$\pm5.8$&$\pm6.9$&$\pm7.8$&$\pm0.58$&$\pm0.92$&$\pm1.2$&$\pm0.21$&$\pm0.24$&$\pm0.26$\\\hline
$\F_{\ell equ\,r|i}^{(1)13}\times10^2$&$\pm2.4$&$\pm2.8$&$\pm3.2$&$\pm0.36$&$\pm0.56$&$\pm0.71$&$\pm0.26$&$\pm0.30$&$\pm0.33$\\\hline
$\F_{\ell equ\,r|i}^{(3)13}\times10^2$ &$\pm0.77$&$\pm0.91$&$\pm1.0$&$\pm0.14$&$\pm0.21$&$\pm0.27$&$\pm0.076$&$\pm0.086$&$\pm0.095$\\\hline
  \end{tabular}
  \caption{Single-coupling bounds on effective four-fermion couplings,
    for three values of $\sqrt{s}$ and three assumed experimental
    uncertainties.} 
  \label{tab:sngl.cplng.4q}
\end{table}

The single-coupling bounds in table \ref{tab:sngl.cplng.4q} show both
a large sensitivity to $\F$ couplings relative to that of $\C$'s,
and a strong enhancement of that sensitivity with increasing energy,
as expected of four-fermion interactions.  As also seen from the
table, the single-top cross section is most sensitive to $\F_{\ell
  q}^{(3)13}$, which is related to the fact that the operator $O_{\ell
  q}^{(3)13}$ in (\ref{eq:operators4q}) is the only four-fermion
operator leading to substantial interference with the SM.  In the case
of $\sqrt{s}=0.5$ TeV, if we assume an experimental uncertainty
$\veps_\mathrm{exp}=30\%$ we obtain the bounds
\begin{equation}
  \label{eq:4q30}
  -0.011<\F_{\ell q}^{(3)13}<0.012,\quad
  \left|\F_{ledq\,r|i}^{13}\right|<0.093,\quad
  \left|\F_{\ell equ\,r|i}^{(1)13}\right|<0.038,\quad
  \left|\F_{\ell equ\,r|i}^{(3)13}\right|<0.012,
\end{equation}
which still are rather strong.

At each energy tables \ref{tab:sngl.cplng} and \ref{tab:sngl.cplng.4q}
give lower and upper bounds for each coupling, for three different
values of $\veps_\mathrm{exp}$.  As is easy to check, any of those
three values results from linear interpolation of the other two,
within about $5\%$.  This shows, heuristically, that linear
interpolation is valid and can be used to find bounds corresponding to
other values of $\veps_\mathrm{exp}$ within the range given in the
table.  Linear extrapolation can also be used to obtain bounds for
$\veps_\mathrm{exp}$ moderately smaller than the lowest value used in
the table, or moderately larger than the highest one.  An illustration
of this is provided by the bounds at $\sqrt{s}=0.5$ TeV and
$\veps_\mathrm{exp}=30\%$ given above in (\ref{eq:bnd30}), (resp.,
(\ref{eq:4q30})) which agree with an extrapolation from table
\ref{tab:sngl.cplng} (resp., \ref{tab:sngl.cplng.4q}) within at most
5\% deviation.

\begin{figure}[th]
  \centering
\includegraphics[scale=1]{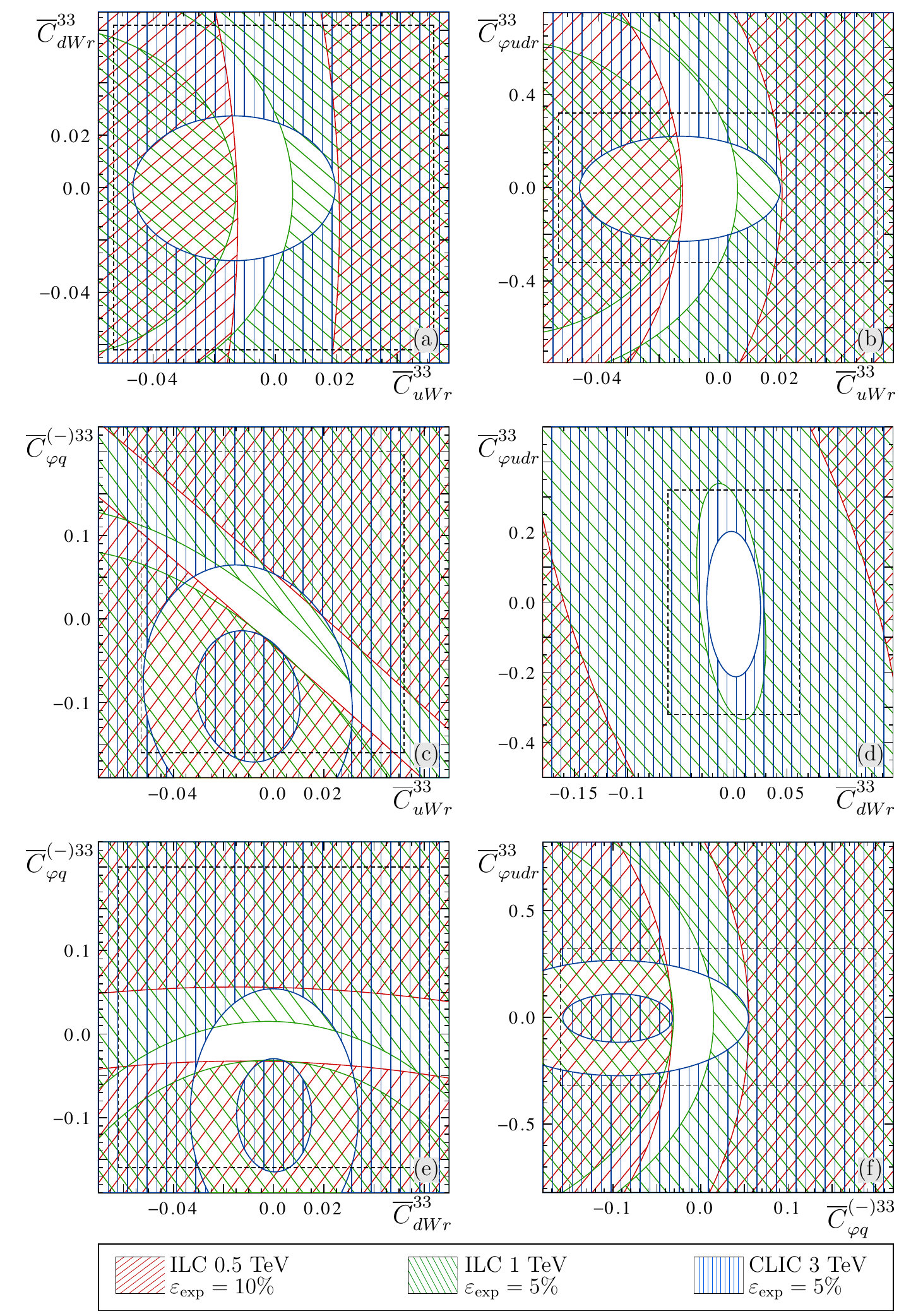}  
\caption{Parameter regions for $tbW$ effective couplings at 68\% C.L.,
  excluded by a measurement of the total cross section for process
  (\ref{eq:prddcy}).  Red hatched areas: $\sqrt{s}=0.5$ TeV, 
  $\varepsilon_\mathrm{exp}=10\%$. Green hatched areas: $\sqrt{s}=1$ TeV,
  $\varepsilon_\mathrm{exp}=5\%$. Blue hatched areas: $\sqrt{s}=3$ TeV,
  $\varepsilon_\mathrm{exp}=5\%$. Dashed lines: CMS bounds from
  (\ref{eq:cmslimfin}).} 
  \label{fig:2cplng}
\end{figure}

\begin{figure}[th]
  \centering
\includegraphics[scale=1.0]{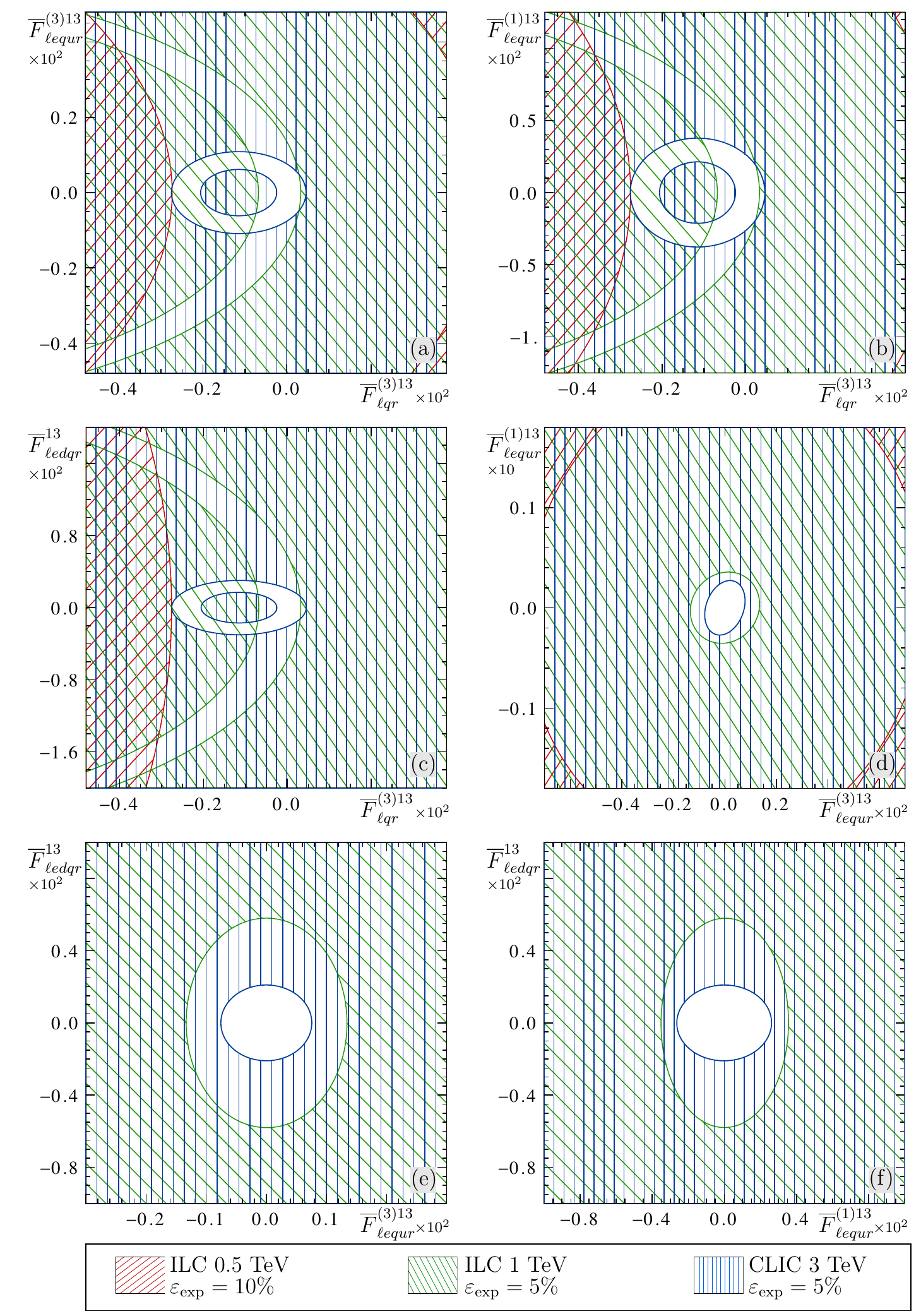}  
\caption{Parameter regions for $tbe\nu_e$ four-fermion charged-current
  couplings at 68\% C.L., excluded by a measurement of the total cross
  section for process (\ref{eq:prddcy}). Color codes as in the previous
  figure.}
  \label{fig:2cplng.4q}
\end{figure}

The allowed regions for pairs of effective couplings involving gauge
bosons are displayed in figure \ref{fig:2cplng}, where the current LHC
bounds from (\ref{eq:cmslimfin}) are also displayed for reference.
The allowed regions for pairs of effective four-fermion couplings are
displayed in figure \ref{fig:2cplng.4q}.  As can be seen in figures
\ref{fig:2cplng}, \ref{fig:2cplng.4q}, with the exception of the
couplings in figures \ref{fig:2cplng}(d) and
\ref{fig:2cplng.4q}(d),(e),(f), the cross section at the ILC at either
$\sqrt{s}=0.5$ or 1 TeV does not determine a small, simply connected
neighborhood of the origin but rather an extended toroidal band.
However, the intersection of those two regions does provide a simply
connected vicinity of the SM.  Those ILC-allowed regions are further
constrained by the bounds imposed by CLIC, as shown in the figures.

The term containing the coefficient $c$ in (\ref{eq:rat2}),
corresponding to the interference of two anomalous amplitudes, leads
to a rotation of the symmetry axes of the allowed region relative to
the coordinate axes.  Interference effects between the amplitudes
proportional to $C_{\vp ud\,r}^{33}$ and $C_{dW\,r}^{33}$ are apparent
in figure \ref{fig:2cplng}(d), sizeable at $\sqrt{s}=0.5$ TeV and
significantly weaker at higher energies.  Smaller, but still
noticeable interference between amplitudes proportional to $\C_{\vp
  q}^{(-)33}$ and $\C_{uW\,r}^{33}$ is seen in figure
\ref{fig:2cplng}(c).  All other pairs of couplings correspond to
effective operators involving $b$-quark fields of opposite
chiralities, for which interference is suppressed by the small
$b$-quark mass.  On the other hand, no interference effects are
visible in figure \ref{fig:2cplng.4q}, except for a weak one in panel
(c) at the lowest $\sqrt{s}$.  In particular, the regions allowed by
the single-top cross section at $\sqrt{s}=0.5$ TeV not shown in
figures \ref{fig:2cplng.4q}(e), (f) are ellipses with their axes
parallel to the coordinate axes, inscribed within the rectangles
defined by the single-coupling bounds in table
\ref{tab:sngl.cplng.4q}.

\begin{figure}[th]
  \centering
\includegraphics[scale=1.0]{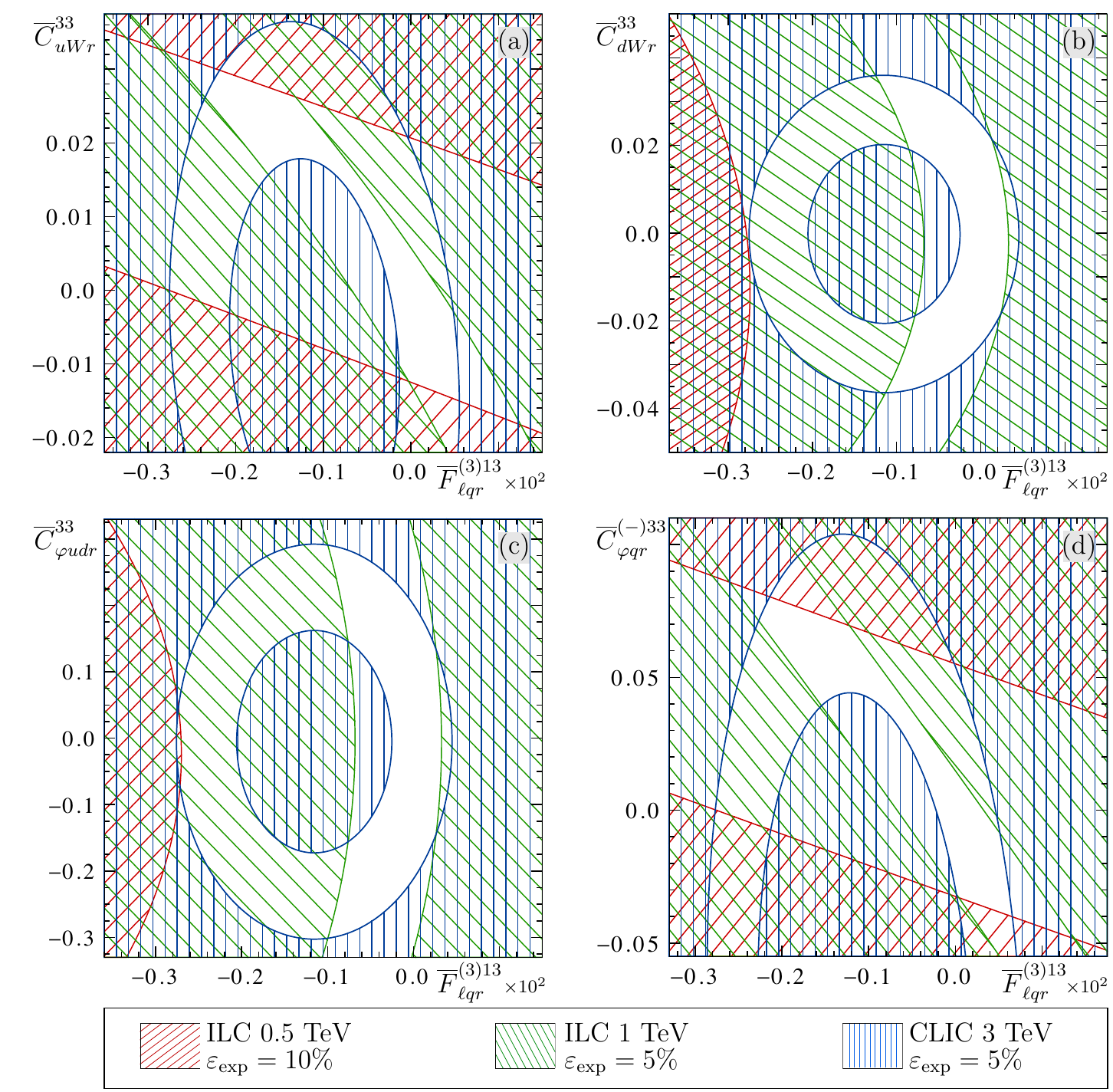}  
\caption{Parameter regions for $tbW$ effective couplings together with
  the $\bar{F}_{\ell qr}^{(3)13}$ four-fermion coupling, excluded at
  68\% C.L. by a measurement of the total cross section for process
  (\ref{eq:prddcy}). Color codes as in the previous figure.}
  \label{fig:2cplng.x}
\end{figure}

As discussed in section \ref{sec:diag4}, the operator $O^{(3)13}_{lq}$
is related to certain gauge-boson operators by the equations of
motion.  In figure \ref{fig:2cplng.x} we show the allowed regions for
pairs of couplings involving $\bar{F}_{\ell qr}^{(3)13}$ and one
gauge-boson effective coupling.

\section{Final remarks}
\label{conclusions}

Top quark physics is an essential part of the future ILC and CLIC
collider programs.  While so far most of the interest has focused on
$t\bar t$ production and its great sensitivity to NC $ttZ(A)$
couplings, in this paper we discuss whether the single-top mode would
provide any useful information.  In this context, it is important to
stress the fact that while at the ILC single-top is
subdominant to top-pair production, which then becomes a strong
background, at CLIC single-top is the dominant top
production mode.  We have carried out a preliminary parton level
analysis of the semi-leptonic six fermion final state $b\bar b e^-
\nu_e q_u q_d$ (and its charge conjugate) in the context of single-top
production and have found that at the three energies considered,
$\sqs=0.5$, 1 and 3 TeV, the signal cross sections are about 1-2 fb
(see table \ref{tab:tab1}), including phase-space cuts and $b$-tagging
efficiencies, enough to keep the statistical uncertainties under 5\%.
A detailed discussion of backgrounds is given in section
\ref{sec:topatee}.

We have obtained individual limits on $tbW$ vertices in the context of
the $SU(2)\times U(1)$-gauge invariant effective dimension-6 operators
\cite{grz10}, as shown in table \ref{tab:sngl.cplng} and figure
\ref{fig:2cplng}.  In order to discuss our results for the four
top-gauge boson couplings generating CC vertices, it is convenient to
separate them in two pairs.  In the first pair we have the operators
$O_{\vp q}^{(-)33}$ and $O_{uW}^{33}$, which have in common that they
both generate interference with the SM, simultaneously generate an
additional NC $ttZ/A$ coupling and contribute substantially to
$s$-channel diagrams that appear in both, single-top and top-pair
production.  With increasing energy both the effects of interference
and the $s$-channel contribution decrease, and so does the sensitivity to
these couplings.
To compare our results to those from the LHC given in
(\ref{eq:cmslimfin}) and to the projections for $t\bar{t}$ production
at the ILC, (\ref{aguilarlimits}) and (\ref{englertlimits}), we take
the inverse of the length of the interval determined by the
single-coupling bounds in those equations and in table
\ref{tab:sngl.cplng} as a measure of the sensitivity.  For the
coupling $\C_{\vp q}^{(-)33}$ the sensitivity obtained at $\sqs=0.5$
TeV if we assume an experimental uncertainty of 10\% is four times
larger than at the LHC, (\ref{eq:cmslimfin}), and about the same as
that of $t\bar{t}$ production at the same energy, with the
experimental uncertainties for $t\bar{t}$ assumed in
(\ref{aguilarlimits}), but three times smaller if the uncertainties
are those assumed in (\ref{englertlimits}).  Larger assumed
experimental uncertainties lead, of course, to correspondingly smaller
sensitivities.  As discussed in relation to (\ref{eq:bnd30}), for an
experimental uncertainty of about 30\%, the sensitivity to $\C_{\vp
  q}^{(-)33}$ at $\sqs=0.5$ TeV becomes equal to the current LHC
result.  As seen from table \ref{tab:sngl.cplng}, the largest
sensitivity to $\C_{\vp q}^{(-)33}$ is obtained at $\sqs=1$ TeV if we
assume an experimental uncertainty of 5\%.  In that case the
sensitivity is about eight times larger than the current LHC result,
almost twice as large as the ILC $t\bar{t}$ sensitivity in
(\ref{aguilarlimits}), and slightly smaller than in
(\ref{englertlimits}).  Thus, for the coupling $\C_{\vp q}^{(-)33}$ we
expect the sensitivity of single-top production at linear $e^-e^+$
colliders to significantly improve the current one at the LHC and the
projected one at the HL-LHC, and to be competitive with the
sensitivity of $t\bar{t}$ production at the ILC.

For the coupling $\C_{uWr}^{33}$, at $\sqs=0.5$ TeV and with an
experimental uncertainty of 10\%, the sensitivity from table
\ref{tab:sngl.cplng} is three times larger than that from the current
LHC result (\ref{eq:cmslimfin}), though almost three times smaller
than the ILC $t\bar{t}$ production result (\ref{aguilarlimits}), and
about 15 times smaller than (\ref{englertlimits}).  The largest
sensitivity to this coupling in table \ref{tab:sngl.cplng} occurs at
$\sqs=1$ TeV for an assumed experimental uncertainty of 5\%, which is
five times larger than the current LHC result (\ref{eq:cmslimfin}),
slightly smaller than the ILC $t\bar{t}$ sensitivity
(\ref{aguilarlimits}) and an order of magnitude smaller than
(\ref{englertlimits}).  In this case we conclude that the sensitivity
to $\C_{uWr}^{33}$ of single-top production at the ILC will improve on
the current LHC sensitivity by a factor 3-5, but will be significantly
worse than that of $t\bar{t}$ production at the ILC.

The operators $O_{\vp ud}^{33}$ and $O_{dW}^{33}$ do not appreciably
contribute to $t\bar{t}$ production at the ILC, and lead to very
modest interference with the SM amplitudes.  They contribute to
single-top production in $e^-e^+$ collisions mostly through
$t$-channel diagrams, with the contribution from $s$-channel diagrams
remaining essentially a SM input that is important at lower energies
but becomes very small for energies higher than $\sqs =2$ TeV.  In
this respect, we notice that $O_{\vp ud}^{33}$ does not generate NC
interactions, and that $O_{dW}^{33}$ only generates, besides $tbW$, a
$bbZ$ coupling with very little effect on $s$-channel diagrams.  As a
result, the sensitivity to these operators increases with energy.

The sensitivity to $\C_{\vp udr}^{33}$ at $\sqs=0.5$ TeV with an
assumed experimental uncertainty of 10\%, as given in table
\ref{tab:sngl.cplng}, is less than half that of the current LHC limits
(\ref{eq:cmslimfin}).  At $\sqs=1$ TeV with an assumed uncertainty of
5\% is equal to the sensitivity of the current LHC results
(\ref{eq:cmslimfin}), and at $\sqs=3$ TeV with uncertainty 5\% is
about 60\% larger than the current LHC sensitivity.  For this
coupling, therefore, at most a slight improvement over the current LHC
sensitivity can be expected from CLIC, and none from the ILC.

The sensitivity to $\C_{dWr}^{33}$ at $\sqs=0.5$ TeV with an assumed
experimental uncertainty of 10\%, as given in table
\ref{tab:sngl.cplng}, is about half that of the current LHC limits
(\ref{eq:cmslimfin}).  At $\sqs=1$ TeV with an assumed uncertainty of
5\% it is twice as large as the current LHC sensitivity, and 2.5 times
as large at $\sqs=3$ TeV with uncertainty 5\%.  Thus, the sensitivity
to $\C_{dWr}^{33}$ at the highest energy $e^-e^+$ collisions increases
by a factor of 2--2.5 with respect to the current LHC sensitivity.

These conclusions depend, of course, on the fact that at each one of
the three energies discussed here, we have assumed the lowest
experimental uncertainties considered in table \ref{tab:sngl.cplng}.
With the information contained in that table, however, these
conclusions can be adapted to other experimental uncertainty
assumptions.

Limits on the four-fermion operators from (\ref{eq:operators4q}) are
given in table \ref{tab:sngl.cplng.4q} and in figures
\ref{fig:2cplng.4q} and \ref{fig:2cplng.x}.  Those CC four-fermion
couplings cannot be observed at hadron colliders, or in NC processes
such as $t\bar{t}$ production at $e^-e^+$ colliders. Therefore,
single-top production at those colliders is the only possibility to
bound them.  Since the sensitivity to four-fermion interactions
increases very rapidly with energy, for these couplings CLIC is
clearly the best option.  The sensitivity at the 1 TeV ILC is still
substantial, however, being about half that for CLIC, as shown in
table \ref{tab:sngl.cplng.4q}.


\paragraph*{Acknowledgments}

We gratefully acknowlegde support from Conacyt of M\'exico through
Research Project 220066, and from Sistema Nacional de Investigadores
de M\'exico.

\end{document}